\documentclass[a4paper,review,11pt,authoryear]{elsarticle}
\usepackage{amsmath,bm,hyperref,url,fullpage,mathtools,booktabs,bbm}
\usepackage{amssymb}
\usepackage{paralist,enumitem,todonotes}
\usepackage{multirow}
\usepackage{threeparttable}
\usepackage{rotating}
\usepackage{float}
\usepackage{graphicx}
\usepackage[ruled,linesnumbered]{algorithm2e}
\usepackage{caption}
\usepackage{subcaption}
\usepackage[a4paper, margin=1in]{geometry}
\usepackage{changes}
\usepackage{enumitem}
\usepackage{algorithm2e}
\usepackage{hyperref}
\hypersetup{
	colorlinks=true,
	linkcolor=blue,
	citecolor=blue,
	urlcolor=blue}
\newtheorem{theorem}{Theorem}  
\newtheorem{proposition}[theorem]{Proposition}

\newtheorem{remark}{Remark}

\newcommand{\x}{\mathbf{x}}
\newcommand{\Nb}{N_b}
\newcommand{\nb}{n_b}
\newcommand{\nt}{n_t}
\newcommand{\D}{\mathcal{D}}
\newcommand{\e}{\epsilon}
\newcommand{\Dcum}{\D_{(b)}}
\newcommand{\bbeta}{\boldsymbol{\beta}}
\newcommand{\truebeta}{\boldsymbol{\beta}_0}
\newcommand{\oraclebeta}{\hat{\boldsymbol{\beta}}^{\star}_b}

\newcommand{\renew}{\tilde{\boldsymbol{\beta}}}

\newcommand{\loss}{\mathcal{L}}

\newcommand{\gradt}{\boldsymbol{g}_t}
\newcommand{\E}{\mathbb{E}}
\newcommand{\renewloss}{\tilde{\loss}}

\newcommand{\ploss}{\breve{\loss}}
\newcommand{\Hess}{\nabla^2\loss}
\newcommand{\pHess}{\nabla^2\ploss}
\newcommand{\sumb}{\sum_{t=1}^{b-1}}

\begin{document}
\date{}
\begin{frontmatter}
  \title{%Online Renewable Expectile Regression for Stream Data with Abnormal Batches
  Renewable Online Expectile Regression for Heterogeneous Streaming Data with Abnormal Batches}
    \author[1]{Wei Cao} 
    \author[1,2]{Shanshan Wang}
    \ead{Corresponding author at (School of Economics and Management, Beihang University, Beijing 100191, China) via sswang@buaa.edu.cn.}
    \address[1]{School of Economics and Management, Beihang
			University, Beijing, China}
	\address[2]{MOE Key Laboratory of Complex System Analysis and Management Decision, Beihang University}
    \begin{abstract}
    
    Streaming data, characterized by high volume, rapid arrival rates, and evolving distributions, have become increasingly prevalent in modern applications. Developing efficient and reliable estimation procedures is therefore essential for real-time statistical analysis. However, most existing online estimation methods rely on the assumption of batch homogeneity, which can be violated in practice due to abnormal batches, distributional shifts, or other forms of batch heterogeneity. To address this challenge, we develop renewable online expectile regression procedures for heterogeneous streaming data. Specifically, we propose two complementary strategies for handling abnormal batches: (1) a detection-based approach that employs a sequential monitoring mechanism based on score test statistics to identify and remove potentially abnormal batches; and (2) an adaptive-weighting approach that assigns data-driven weights to incoming batches, reducing the influence of abnormal or drifting batches while retaining information from reliable observations. Both strategies rely solely on score test statistics and can be seamlessly integrated into existing renewable estimation and inference frameworks without requiring additional structural assumptions. Furthermore, to enhance robustness against heavy-tailed errors and outliers, we replace the conventional $\ell_2$ loss with the Huber loss and develop a robust extension of renewable online expectile regression. Extensive simulation studies and analyses of clinical datasets demonstrate that the proposed methods achieve improved estimation accuracy and robustness in the presence of batch heterogeneity. Overall, the proposed framework provides a flexible and effective solution for renewable expectile regression in complex streaming data environments.

    \end{abstract}
				
    \begin{keyword}
    Streaming data \sep Expectile regression \sep Renewable estimation \sep Robust Regression
    \end{keyword}
		
\end{frontmatter}
	
\section{Introduction}
\label{sec:introduction}

The analysis of large-scale streaming data has attracted increasing attention, as such data are continuously generated at high velocity and often must be processed under storage and computational constraints \citep{gama2013evaluating, fan2014challenges}. Online updating provides a natural solution to this setting by enabling sequential incorporation of newly arriving data without repeatedly accessing historical observations. Various online estimation strategies have been developed, including cumulatively updated estimating equations \citep{lin2011aggregated, schifano2016online}, stochastic gradient descent and its variants \citep{chen2020statistical, Zhu02012023}, and renewable estimators \citep{luo2020renewable}, which achieve computational efficiency by maintaining only low-dimensional summary information from previous data batches.

Beyond their sequential nature, streaming data often exhibit distributional shifts across batches or may contain a small number of anomalous batches. Such phenomena are frequently observed in clinical applications, where data distributions can evolve over time due to institutional differences in staffing, instrumentation, and data-collection workflows; changes in clinical practices; or variations in patient characteristics, including race and ethnicity, sex, age, and socioeconomic status \citep{subbaswamy2021evaluating, subasri2025detecting}. Ignoring such batch-wise distributional changes may result in biased estimation and misleading scientific conclusions. 

To address this issue, recent studies have incorporated batch-level heterogeneity into streaming-data analysis from various perspectives. For example, \citet{lu2021online} proposed homogenized representations to mitigate distributional discrepancies across data batches; \citet{wei2023adaptive}, \citet{luo2023multivariate}, and \citet{liu2025online} developed adaptive or dynamic updating strategies that allow model parameters to evolve with incoming data; and \citet{ding2024renewable} and \citet{chen_renewable_2024} introduced penalized renewable estimation procedures to accommodate distributional drift in online inference. More recently, \citet{sheng2024sequential} proposed a shift-adjusted sequential integration framework based on pairwise pseudo-likelihood to eliminate unknown shift functions. These approaches provide valuable tools for handling gradual distributional drift and batch-level heterogeneity in streaming environments. However, in some applications, heterogeneity may be primarily caused by a small number of abnormal batches, representing a localized and abrupt departure from the underlying data-generating mechanism. Such settings require targeted procedures that can identify and appropriately handle abnormal batches before their information is propagated through subsequent online updating.

To address potentially abnormal or contaminated data batches, several studies have developed monitoring procedures to identify incompatible batches and prevent them from adversely affecting subsequent model updates. For example, \citet{luo2023real} proposed an online framework equipped with a goodness-of-fit monitoring procedure to detect abnormal or incompatible batches, such that model updating is conducted only when newly arriving data are sufficiently consistent with previously accumulated information. Similar monitoring strategies have been further extended to renewable quantile regression and generalized estimating equations \citep{wang2025real, wang2026renewable}. 
Despite these advances, several challenges remain to be addressed. First, detection-based approaches typically rely on pre-specified thresholds to determine whether a newly arriving batch should be incorporated. Although such thresholding rules provide a simple and effective mechanism for online screening, their performance may be sensitive to threshold selection, particularly when the discrepancy between the incoming batch and the historical model is moderate rather than substantial \citep{gunasekaran2025predictive}. Second, existing detection-based approaches generally follow an ``accept-or-reject’’ updating strategy, where all observations within a potentially abnormal batch are either fully incorporated or completely discarded. While this strategy effectively protects online updates from severely incompatible batches, it may be suboptimal in more nuanced scenarios. Specifically, failing to detect a harmful batch may introduce contaminated information into the updating process and compromise model accuracy, whereas discarding an entire batch may unnecessarily remove informative observations from partially compatible batches. This issue is closely related to recent developments in transfer learning, where identifying and exploiting partially transferable information has become increasingly important \citep{zhao2026residual}.

The aforementioned monitoring procedures mainly focus on batch-level compatibility, where an entire incoming batch is classified as abnormal or incompatible with historical information. In many streaming applications, however, heterogeneity and contamination may also occur within otherwise informative batches. Such within-batch irregularities can arise from heteroscedastic errors, heterogeneous covariate effects, asymmetric error distributions, or sample-level contamination. Quantile regression (QR) and expectile regression (ER) provide two important frameworks for characterizing such distributional heterogeneity. For QR, various online updating methods have been developed based on one-shot estimators, likelihood-based approaches, and sequential Bayesian algorithms \citep{chen2020quantile, wang_renewable_2022, fanSequentialQuantileRegression2024a}, with recent extensions to high-dimensional settings \citep{jiangRenewableQuantileRegression2022, xieStatisticalInferenceSmoothed2024, sunOnlineRenewableSmooth2023, peng2024two}. 

Compared with QR, ER is formulated through asymmetric least squares and offers a computationally efficient alternative for modeling heterogeneous covariate effects and asymmetric conditional distributions \citep{newey1987asymmetric}. Consequently, online ER has received increasing attention in recent years \citep{song2021linear, pan2024renewable, cao2026renewable}. However, the quadratic nature of the standard ER loss makes it vulnerable to outliers and heavy-tailed observations, which are frequently encountered in streaming environments. To improve robustness, several robust ER approaches have been proposed, such as replacing the conventional loss function with the Huber loss \citep{man2024retire} or introducing separate tuning parameters to simultaneously regulate asymmetry and robustness \citep{zhao2022robust}. Nevertheless, existing robust ER methods remain largely developed for static or conventional settings, and robust online renewable ER methods that can simultaneously accommodate batch-level heterogeneity and sample-level contamination are still underexplored.

Motivated by these limitations, this paper develops a robust renewable expectile regression framework for streaming data subject to batch-level incompatibility and sample-level contamination. The proposed framework consists of three integrated components. First, we construct a Lagrange multiplier-type diagnostic statistic for monitoring newly arrived batches. This formal testing procedure provides an efficient screening mechanism for identifying batches that are strongly incompatible with the accumulated historical information. Second, recognizing that hard detection may be suboptimal when batch heterogeneity is moderate, we propose an adaptive weighting strategy based on a guided loss function. The adaptive weights dynamically balance historical information and newly arrived data, enabling the updating procedure to exploit partially compatible batches while mitigating the impact of less compatible batches. Third, to enhance robustness against outliers and heavy-tailed errors within incoming batches, we incorporate the Huber loss \citep{huber1964robust} into the renewable ER framework and develop a robust online updating procedure. Extensive numerical studies and empirical data analyses further demonstrate the effectiveness of the proposed method.

The remainder of the article is organized as follows. Section~\ref{sec:preliminary} briefly reviews the renewable estimation procedure under a homogeneous data setting and formulates the problem considered in this article. Sections~\ref{sec:DReER} and~\ref{sec:AReER} develop two strategies for accommodating streaming data containing potentially abnormal batches and present the corresponding implementation algorithms.  For simplicity, we refer to these strategies as (1) detection-based renewable ER (DReER) and (2) adaptive-weighting renewable ER (AReER), respectively. Section~\ref{sec:robust} extends the proposed framework to a robust Huber-type formulation. Section~\ref{sec:experiments} investigates the finite-sample performance of the proposed methods through comprehensive simulation studies. Section~\ref{sec:empirical} presents an application to a real-world dataset. Finally, Section~\ref{sec:conclusion} summarizes the main findings and discusses potential directions for future research. Technical details and additional numerical results are provided in the Appendix.

\section{Preliminaries and Problem Formulation}\label{sec:preliminary}

For any $\tau\in(0,1)$, let $e_\tau(Y|X=\x)$ denote the conditional expectile of $Y$ given $X=\x$ at $\tau$-th expectile, which can be obtained by minimizing the asymmetric squared loss function \citep{newey1987asymmetric}:
\begin{equation*}
	e_\tau(Y|X=\x) = \mathop{\arg\min}\limits_{\theta} \E\Big(\rho_\tau(Y-\theta)|X=\x\Big),
\end{equation*}
where $\rho_\tau(u)=\frac{1}{2}\cdot u^2\cdot|\tau-\mathbf{I}(u<0)|$, and $\mathbf{I}(u<0)$ is an indicator function. 

Suppose that $N_b$ samples are sequentially generated in $b$ data batches, denoted by $\{\mathcal{D}_1,\ldots,\mathcal{D}_b\}$, where $\mathcal{D}_t=\{(\mathbf{x}_{ti},y_{ti})\}_{i=1}^{n_t}$ represents the $n_t$ samples in the $t$-th batch. Thus, $N_b=\sum_{t=1}^b n_t$, and $\mathcal{D}_{(b)}$ denotes the streaming dataset accumulated up to and including batch $b$. For each batch $\D_t$, sample $(\x_{ti},y_{ti})$ follows the linear expectile regression model:
\begin{equation}
\label{eq:expectile}
    e_\tau(y_{ti}|\x_{ti})=\x_{ti}^\top \truebeta(\tau), \quad i=1,\ldots, n_t,
\end{equation}
where $\x_{ti}=(x_{ti,1},\ldots,x_{ti,p})^\top$ is a $p \times 1$  covariate vector. Here, $\truebeta (\tau)$ is the unknown expectile regression parameter. To simplify notation, we omit  $\tau$  where there is no ambiguity. 

Suppose $\Nb$ samples are available, 
the conventional offline estimator $\oraclebeta$ could be obtained by minimizing the following global loss function: 
\begin{equation*}
\begin{aligned}
\oraclebeta=\mathop{\arg\min}\limits_{\bbeta} \ \loss_{\Nb}(\bbeta)=\mathop{\arg\min}\limits_{\bbeta}\left[\frac{1}{N_b}\sum_{t=1}^b\sum_{i=1}^{n_t}\rho_\tau\left(y_{ti}-\x_{ti}^\top\bbeta\right)\right],
\end{aligned}
\end{equation*}
which can be solved using an iteratively reweighted least squares (IRLS) method. However, in a streaming setting, the cumulative dataset $\Dcum$ cannot be accessed all at once. 

To address this issue, \cite{cao2026renewable} proposed a renewable estimation procedure that utilizes the current data batch together with information from historical raw data summarized through summary statistics. Specifically, 
let $\loss_{\nt}(\bbeta)$ and $\loss_{N_b}(\beta)$ denote the batch level loss of $\D_t$ and  cumulative global loss up to batch $b$, respectively. 
Thus, 
\begin{equation*}
\label{eq:loss-cumbatch}
\begin{aligned}
\loss_{N_b}(\beta)=\frac{1}{N_b}\sum_{t=1}^bn_t\cdot\loss_{\nt}(\bbeta),\qquad \loss_{\nt}(\bbeta)=\frac{1}{n_t}\sum_{i=1}^{n_t}\rho_\tau\left(y_{ti}-\x_{ti}^\top\bbeta\right).
\end{aligned}
\end{equation*}
Let $\renew_{b-1}$ denote the renewable estimator obtained using the data up to batch $b-1$. When $\D_b$ arrives, the overall loss function $\loss_{N_b}(\bbeta)$ can be decomposed into contributions from the historical and current batches:
\begin{equation}\label{eq:loss-decompose}
    \loss_{N_b}(\bbeta)=\frac{1}{\Nb}\Big(N_{b-1}\loss_{N_{b-1}}(\bbeta)+\nb\loss_{\nb}(\bbeta)\Big).
\end{equation}
By applying a Taylor expansion around the previous renewable estimator $\renew_{b-1}$ on the historical term of \eqref{eq:loss-decompose}, the cumulative loss $\loss_{N_{b}}(\bbeta)$ can be approximately represented as:
\begin{equation}\label{eq:online-taylor}
    \tilde\loss_{N_{b}}(\bbeta)=\frac{1}{\Nb}\Bigg\{\frac{1}{2}(\bbeta-\renew_{b-1})^\top\Bigg[\sumb\nt \Hess_{\nt}(\renew_{t})\Bigg](\bbeta-\renew_{b-1})+n_b\loss_{\nb}(\bbeta)\Bigg\},
\end{equation}
where $\Hess_{\nt}(\renew_{t})$ is the Hessian of the loss function of data batch $t$ evaluated at $\renew_{t}$. Minimizing Eq.\eqref{eq:online-taylor} yields the renewable estimator $\renew_b$ satisfying
\begin{equation}\label{eq:online-sol}
\renew_b = 
\left[ \sumb \mathbf{W}_{t}(\renew_t) + \mathbf{W}_{b}(\renew_b) \right]^{-1} 
\left[ \sumb \mathbf{W}_{t}(\renew_t) \renew_{b-1} + \mathbf{U}_{b}(\renew_b) \right],
\end{equation}
where $\mathbf{W}_{t}(\renew_t)$ and $\mathbf{U}_{t}(\renew_t)$ are defined as:
\begin{equation*}
    \mathbf{W}_{t}(\renew_t)=\sum_{i=1}^{\nt} \left|\tau-\mathbf{I}(y_{ti}<\x_{ti}^{\top}\renew_t)\right|\x_{ti}\x_{ti}^\top,\quad 
    \mathbf{U}_{t}(\renew_t)=\sum_{i=1}^{\nt} \left|\tau-\mathbf{I}(y_{ti}<\x_{ti}^{\top}\renew_t)\right|\x_{ti}y_{ti}.
\end{equation*}
Since the right-hand side of Eq.\eqref{eq:online-sol} still involves the unknown parameter $\renew_b$, we replace $\renew_b$ with the previous estimator $\renew_{b-1}$ to improve computational efficiency. The resulting renewable estimator is then given by:
\begin{equation}\label{eq:renewable}
   \renew_b = 
    \left[ \sumb \mathbf{W}_{t}(\renew_t) + \mathbf{W}_{b}(\renew_{b-1}) \right]^{-1} 
    \left[ \sumb \mathbf{W}_{t}(\renew_t) \renew_{b-1} + \mathbf{U}_{b}(\renew_{b-1}) \right].
\end{equation}

\cite{cao2026renewable} established the consistency and asymptotic normality of $\renew_b$ under mild regularity conditions and showed that it achieves the same statistical efficiency as the oracle estimator $\oraclebeta$ based on the full data. However, the aforementioned renewable estimation procedure is designed for homogeneous streaming data, where all data batches are assumed to share the same regression parameter. Under this homogeneity assumption, the corresponding data-generating mechanism is referred to as the reference model. In practice, this assumption may be violated due to structural changes, abnormal batches, or distributional shifts. 

To address this issue, we next develop a renewable expectile regression method that is robust to abnormal batches. Here, let $\iota \in \{1,2,\ldots,b\}$ denote the index of a potentially abnormal batch. In the presence of heterogeneity across data batches, the underlying regression parameter of batch $\mathcal{D}_{\iota}$ may differ from that of the reference model. Specifically, we allow $\bbeta{\iota} \neq \truebeta$, indicating that $\mathcal{D}_{\iota}$ is generated from a model that is incompatible with the reference model~\eqref{eq:expectile}. If such a batch is naively incorporated into the renewable updating procedure, the resulting cumulative estimator may accumulate systematic bias and gradually deviate from $\truebeta$.

To mitigate the impact of such batch-level structural shifts, we propose two strategies, presented in Sections~\ref{sec:DReER} and~\ref{sec:AReER}, respectively. The first is a detection-based approach that monitors each incoming batch and excludes those identified as potentially abnormal. The second is an adaptive weighting approach that assigns lower weights to batches that may deviate from the reference model.

\section{Detection-Based Renewable Expectile Regression with Abnormal Data Batches}\label{sec:DReER}

%\subsubsection{Detective Monitor and Model Framework}\label{sec2-sub1-1}

\subsection{DReER method}\label{sub:dreer}

The first strategy aims to monitor each incoming batch and detect potential structural deviations before incorporating it into the renewable updating procedure. To this end, we adopt a Lagrange multiplier (LM)-type testing procedure based on the batch-specific score function. Let $\gradt(\bbeta)$ denote the score function associated with the batch-specific loss function $\loss_{\nt}(\bbeta)$, defined as follows: 
\begin{equation*}\label{eq:score}
\gradt(\bbeta)
=
\frac{\partial \loss_{\nt}(\bbeta)}{\partial \bbeta}
=
-\frac{1}{\nt}
\sum_{i=1}^{\nt}
|\tau-\mathbf{I}(y_{ti}<\x_{ti}^{\top}\bbeta)|
\x_{ti}
(y_{ti}-\x_{ti}^{\top}\bbeta).
\end{equation*}
Under the homogeneity assumption $H_0:\bbeta_t=\bbeta_0$, the score function satisfies the first-order optimality condition
$\mathbb{E}\left[\gradt(\truebeta)\right]=\mathbf{0}$.

Let $\epsilon_{ti}=y_{ti}-\x_{ti}^{\top}\bbeta$ denote the model error and define
$\psi_{ti}(\bbeta)
=
\left|
\tau-\mathbf{I}(\epsilon_{ti}<0)
\right|\x_{ti}\epsilon_{ti}$,
which is the gradient vector of the loss function with respect to observation $i$ in batch $t$. Furthermore, let
$\epsilon_{ti}^{0}
=
y_{ti}-\x_{ti}^{\top}\bbeta_0$
denote the error under the homogeneity assumption. Define the covariance matrix of score function as:
$$\mathbf{C}_t
=
\mathrm{Cov}\bigl(\psi_{ti}(\bbeta_0)\bigr)
=
\mathbb{E}
\left[
\left|
\tau-\mathbf{I}(\epsilon_{ti}^{0}<0)
\right|^{2}
(\epsilon_{ti}^{0})^{2}
\x_{ti}\x_{ti}^{\top}
\right].$$
In addition we introduce some regularity condition:
\begin{enumerate}[label=(C\arabic*), topsep=0pt, itemsep=0pt, parsep=0pt]
    \item \label{con1}
    Under null assumption, $\mathbb{E}[\psi_{ti}(\boldsymbol{\beta}_0)] = \mathbf{0}$.
    \item \label{con2}
    The gradient vector has finite second moment,
    $\mathbb{E}\|\psi_{ti}(\boldsymbol{\beta}_0)\|_2^2 < \infty$.
    \item \label{con3}
    $\mathbf{C}_t$ is a positive definite matrix.
\end{enumerate}

Then, under these regularity conditions, the central limit theorem yields $ \sqrt{\nt}\gradt(\bbeta_0)
    \xrightarrow{d}
    N\left(\mathbf{0}, \mathbf{C}_t\right)$. 
Consequently, the LM test statistic is constructed as
\begin{equation}\label{eq:LM-test}
    \Lambda_t(\bbeta_0)=n_t \gradt(\bbeta_0)^{\top}
\widehat{\mathbf{C}}_t^{-1} \gradt(\bbeta_0),
\end{equation}
where $\widehat{\mathbf{C}}_t=\frac{1}{n_t}\sum_{i=1}^{n_t}\left[\left|\tau-\mathbf{I}(\epsilon_{ti}^{0}<0)\right|^2 \left(\epsilon_{ti}^{0}\right)^{2} \boldsymbol{x}_{ti}\boldsymbol{x}_{ti}^\top\right]$. 
Under the null hypothesis of homogeneity and the regularity conditions \ref{con1}-\ref{con3}, the LM test statistic asymptotically follows $\chi^2$ distribution with degrees of freedom $p$, i.e., 
$\Lambda_t(\bbeta_0)\xrightarrow{d}\chi_p^2$,
as established in \cite{newey1987asymmetric}. The LM test statistic allows us to assess whether the incoming data batch $\mathcal{D}_t$ is compatible with the reference model. If $H_0$ is rejected, indicating that $\mathcal{D}_t$ is inconsistent with the reference model, we do not use $\mathcal{D}_t$ to update $\renew_{t-1}$ and set $\renew_t=\renew_{t-1}$. Otherwise, $\renew_{t-1}$ is updated to $\renew_t$ using the procedure proposed in Section \ref{sec:preliminary}. We then proceed to test the homogeneity hypothesis using the next data batch $\mathcal{D}_{t+1}$.

\begin{figure}[htbp]
    \centering
    \includegraphics[width=1\linewidth]{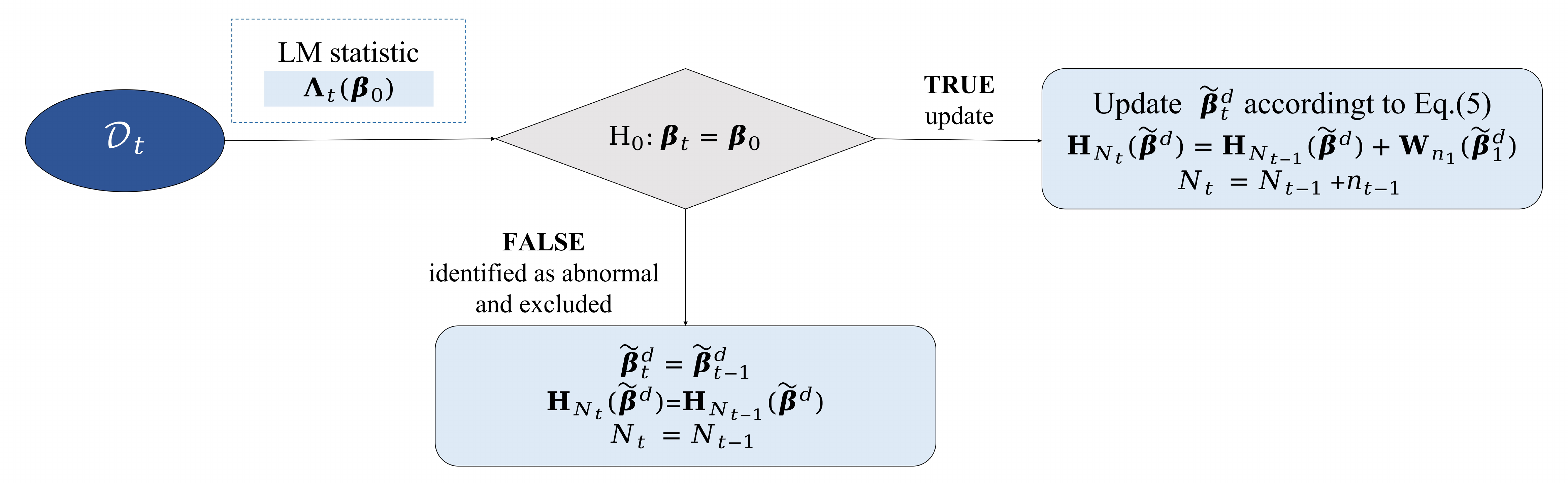}
    \caption{Flowchart of DReER model for abnormal data batch.}
    \label{fig:det}
\end{figure}

The framework of the detection-based renewable expectile regression (DReER) is illustrated in Figure~\ref{fig:det}. For each newly arriving data batch, we first compute the LM statistic to test the null hypothesis that the incoming data are generated from the reference model. If the null hypothesis is rejected, the current batch is identified as abnormal and excluded from the online updating procedure, while the parameter estimates and historical statistics are retained. Otherwise, if the LM test fails to reject the null hypothesis, we update both the parameter estimates and the cumulative historical statistics using the newly arriving batch. For convenience, we define the detection-based renewable estimator as $\renew_t^d$ and the cumulative historical statistics as $\mathbf{H}_{N_t}(\renew^d)=\sum_{j=1}^t\mathbf{W}_{n_t}(\renew_t^d)$, where $\tilde{\boldsymbol{\beta}}^d=\{\tilde{\boldsymbol{\beta}}^d_1,\ldots,\tilde{\boldsymbol{\beta}}^d_t\}$.

\begin{remark}\label{remark-1}

Under classical statistical inference, the likelihood ratio (LR), Wald, and LM tests are asymptotically equivalent under the null hypothesis and under local alternatives. Thus, any of these tests can, in principle, be used to detect departures from the reference model. Nevertheless, they differ substantially in their implementation and computational requirements. The LR test requires estimation under both the restricted (null) and unrestricted models, whereas the Wald test relies on the unrestricted estimator. In contrast, the LM test only requires estimation under the null hypothesis and evaluates the score function at the restricted estimator.

This distinction is particularly important in the streaming-data setting, where data batches arrive sequentially and potentially at high frequency, while the sample size of each batch may be relatively small. Repeatedly fitting unrestricted models for every incoming batch can impose substantial computational costs and undermine the efficiency of online monitoring. Therefore, although the LR and Wald tests are also valid alternatives, we adopt the LM test primarily for its computational convenience. Since it only requires the restricted estimator and the corresponding score function, the LM test avoids repeated estimation of unrestricted models and is thus well suited to online, batch-wise monitoring and renewable estimation procedures.
\end{remark}

\subsection{Implementation of DReER}\label{sub:implementation-dreer}

Note that the LM test statistic defined in \eqref{eq:LM-test} depends on the true parameter $\truebeta$ under the null hypothesis, which is unknown in practice. Following the plug-in strategy adopted in the existing literature, we use the estimator obtained from the first data batch as a plug-in estimate of $\truebeta$. Under the homogeneity assumption, the first data batch is generated from the reference model. By the consistency of the ER estimator, the first-batch estimator converges in probability to $\truebeta$ as $n_1\to\infty$. Hence, the first-batch estimator can be used as a consistent plug-in estimate of $\truebeta$ when constructing the LM test statistic for subsequent data batches.

Specifically, let $\hat{\bbeta}_0=\hat{\bbeta}_1$. For a newly arriving data batch $\mathcal{D}_t$, define $\hat{\epsilon}_{ti}^{0} = y_{ti} - \boldsymbol{x}_{ti}^\top \hat{\boldsymbol{\beta}}_0$. The local score function and covariance matrix are then defined as follows:
\begin{equation*}
\begin{aligned}
\hat{\boldsymbol{g}}_t(\hat{\boldsymbol{\beta}}_0) 
&=
-\frac{1}{n_t}\sum_{i=1}^{n_t}
\left|\tau - \mathbf{I}\left(\hat{\epsilon}_{ti}^{0}<0\right)\right| 
\cdot\boldsymbol{x}_{ti}\hat{u}_{ti}, \\
\widehat{\mathbf{C}}_t(\hat{\boldsymbol{\beta}}_0)
&=
\frac{1}{n_t}\sum_{i=1}^{n_t}
\left[
\left|\tau - \mathbf{I}\left(\hat{\epsilon}_{ti}^{0}<0\right)\right| ^2
\cdot \left(\hat{\epsilon}_{ti}^{0}\right)^2
\cdot \boldsymbol{x}_{ti}\boldsymbol{x}_{ti}^\top
\right].
\end{aligned}
\end{equation*}
Hence, the LM test statistic of data batch $\mathcal{D}_t$ is
\begin{equation} \label{eq:LM-est}
\widehat{\Lambda}_t(\hat{\boldsymbol{\beta}}_0)
=
n_t\,
\hat{\boldsymbol{g}}_t(\hat{\boldsymbol{\beta}}_0)^\top
\left(\widehat{\mathbf{C}}_t(\hat{\boldsymbol{\beta}}_0)\right)^{-1}
\hat{\boldsymbol{g}}_t(\hat{\boldsymbol{\beta}}_0).
\end{equation}

Here, Eq.\eqref{eq:LM-est} is used to assess whether the newly arriving data batch departs from the reference model. In practice, for a significance level $\alpha>0$, let $\chi^2_{p,1-\alpha}$ denote the $(1-\alpha)$-quantile of the $\chi^2_p$ distribution. If $\widehat{\Lambda}_t(\hat{\boldsymbol{\beta}}_0) > \chi^2_{p,1-\alpha}$, the null hypothesis is rejected at the significance level $\alpha$, and the $t$-th batch is identified as abnormal and excluded from the renewable updating procedure. Otherwise, $\mathcal{D}_t$ is regarded as compatible with the reference model and incorporated into the cumulative estimator. This detection step ensures that only data batches consistent with the reference model contribute to the renewable estimator.

Finally, Algorithm~\ref{alg:dreer} summarizes the implementation of DReER. Accordingly, the corresponding quantities are redefined as
\begin{align*}
%\label{eq:renewable-det}
   \renew_b^d = 
    \left[ \sumb \mathbf{W}_{t}(\renew_t^d) + \mathbf{W}_{b}(\renew_{b-1}^d) \right]^{-1} 
    \left[ \sumb \mathbf{W}_{t}(\renew_t^d) \renew_{b-1}^d + \mathbf{U}_{b}(\renew_{b-1}^d) \right],
\end{align*}
where $\mathbf{W}_{n_{t}}(\tilde{\boldsymbol{\beta}}_{t}^d)$ and $\mathbf{U}_{n_{t}}(\tilde{\boldsymbol{\beta}}_{t}^d)$ are define as
\begin{equation*}
    \mathbf{W}_{{t}}(\tilde{\boldsymbol{\beta}}_{t}^d)
    =\sum_{i=1}^{n_t}\left|\tau
    - \mathbf{I}(y_{ti}<\boldsymbol{x}_{ti}^{\top}\tilde{\boldsymbol{\beta}}_{t}^d)\right|
    \boldsymbol{x}_{ti}\boldsymbol{x}_{ti}^{\top}, \quad
    \mathbf{U}_{{t}}(\tilde{\boldsymbol{\beta}}_{t}^d)
    =\sum_{i=1}^{n_t}\left|\tau
    - \mathbf{I}(y_{ti}<\boldsymbol{x}_{ti}^{\top}\tilde{\boldsymbol{\beta}}_{t}^d)\right|
    \boldsymbol{x}_{ti}y_{ti}.
\end{equation*}
Overall, under the DReER framework, after each data batch arrives, we only need to update and store the summary statistics $\mathcal{S}_t = \{\mathbf{H}_{N_t}(\tilde{\boldsymbol{\beta}}^d),\tilde{\boldsymbol{\beta}}^d,N_t\}$. Thus, the DReER framework requires only the updated summary statistics $\mathcal{S}_t$ to be stored after each batch arrival, without retaining the historical raw data.

\begin{algorithm}[H]
    \linespread{.95}\selectfont
    \caption{DReER: Detection-Based Renewable Expectile Regression}
    \label{alg:dreer}
     \SetKwData{Left}{left}\SetKwData{This}{this}\SetKwData{Up}{up}
    \SetKwFunction{Union}{Union}\SetKwFunction{FindCompress}{FindCompress}
    \SetKwInOut{Input}{Input}\SetKwInOut{Output}{Output}

    \Input{Streaming data $\mathcal{D}_{(b)}=\{\mathcal{D}_t\}_{t=1}^b$ with $\mathcal{D}_t=\{\boldsymbol{x}_{ti},y_{ti}\}_{i=1}^{n_t}$, expectile level $\tau$.}
    \Output{Renewable estimation under $\tau$ level $\tilde{\boldsymbol{\beta}}^d_t(\tau)$ and summary statistics $\mathcal{S}_t$ for $t=1,\dots,b$}

\textbf{Initialization:}\\
Compute the initial estimator $\hat{\boldsymbol{\beta}}_1$ using $\D_t$, and set $\hat{\boldsymbol{\beta}}_1=\hat{\boldsymbol{\beta}}_0=\renew_1^d$;\\
Compute and store summary statistics:
\begin{align*}
\mathcal{S}_1&=\{\mathbf{H}_{N_1}(\tilde{\boldsymbol{\beta}}^d),
\tilde{\boldsymbol{\beta}}^d_1,
N_1=n_1\},\\
\mathbf{H}_{N_1}(\tilde{\boldsymbol{\beta}}^d)
&=\mathbf{W}_{1}(\tilde{\boldsymbol{\beta}}^d_1)=
\sum_{i=1}^{n_1}|\tau-\mathbf{I}(y_{1i}<\boldsymbol{x}_{1i}^\top
\tilde{\boldsymbol{\beta}}^d_1)|
\boldsymbol{x}_{1i}\boldsymbol{x}_{1i}^\top.
\end{align*}
\For{$t=2$ \KwTo $b$}{
    \textbf{Step 1: Compute the detection statistic }\\
    Compute the LM test statistic for batch $t$, denoted by $\widehat{\Lambda}_t(\hat{\boldsymbol{\beta}}_0)$.

    \textbf{Step 2: Abnormal detection and online updating}\\
    \eIf{$\widehat{\Lambda}_t(\hat{\boldsymbol{\beta}}_0) > \chi^2_{p,1-\alpha}$}{
        \textbf{Reject}\\
        Treat data batch $t$ as abnormal and remove it from the online updating procedure.\\
        Retain the previous estimator and cumulative statistics:
        \[
        \tilde{\boldsymbol{\beta}}^d_t = \tilde{\boldsymbol{\beta}}^d_{t-1}, \quad N_t = N_{t-1}, \quad \mathbf{H}_{N_t}(\tilde{\boldsymbol{\beta}}^d) = \mathbf{H}_{N_{t-1}}(\tilde{\boldsymbol{\beta}}^d).
        \]
    }{
        \textbf{Accept}\\
        Update the renewable estimator $\tilde{\boldsymbol{\beta}}^d_t$.\\
        Update the cumulative statistics (sample size and accumulated Hessian matrix):
        \[
        N_t = N_{t-1} + n_t, \quad
        \mathbf{H}_{N_t}(\tilde{\boldsymbol{\beta}}^d)
        = \mathbf{H}_{N_{t-1}}(\tilde{\boldsymbol{\beta}}^d)
        +  \mathbf{W}_{t}(\tilde{\boldsymbol{\beta}}^d_t).
        \]
    }

    \textbf{Step 3: Store historical statistics  $\mathcal{S}_t = \{\mathbf{H}_{N_t}(\tilde{\boldsymbol{\beta}}^d), 
    \tilde{\boldsymbol{\beta}}^d_t, N_t\}$ and discard data batch $\mathcal{D}_t$.}\\
}
\end{algorithm}

\section{Adaptive Weighting Renewable Expectile Regression with Abnormal Batches}\label{sec:AReER}

In Section~\ref{sec:DReER}, we introduced a detection-based procedure to monitor whether a newly arriving batch exhibits heterogeneity relative to the reference model. However, as discussed in Section~\ref{sec:introduction}, heterogeneity across batches can arise in various forms. In addition to abrupt structural changes, gradual distributional drift across batches may accumulate over time and eventually lead to biased estimation. In such settings, the performance of the detection-based approach may be unsatisfactory, as its effectiveness depends critically on the statistical power of the underlying hypothesis test and the choice of the detection threshold.

When the discrepancy between batches is relatively mild, the test may fail to reject the null hypothesis despite the presence of small but systematic deviations. Furthermore, different choices of the detection threshold can yield inconsistent detection results, thereby compromising the stability and robustness of the renewable updating procedure. These limitations suggest that a purely detection-based mechanism may be inadequate for handling smoothly evolving heterogeneity in streaming environments. 

\subsection{AReER method}\label{sub:areer}

To address this issue, inspired by the idea of \citet{gunasekaran2025predictive}, we introduce an adaptive weighting mechanism to stabilize the renewable updating process when an incoming batch deviates from the reference model. Rather than discarding potentially heterogeneous batches, the proposed strategy continuously adjusts the contribution of each incoming batch through a weighted loss formulation.

Specifically, we construct an adaptive batch-specific loss function by introducing a weight parameter $\gamma_t \in [0,1]$ to control the contribution of the newly arrived batch. The adaptive loss function for batch $t$ is defined as follows: 
\begin{equation*}
\loss_{n_t}^a(\bbeta)
=\gamma_t\cdot\loss_{n_t}(\bbeta)
+(1-\gamma_t)\cdot\ploss_{n_t}(\bbeta),
\end{equation*}
where $\ploss_{n_t}(\bbeta)=\frac{1}{\nt}\sum_{i=1}^{\nt}\rho_{\tau}(\breve{y}_{ti}-\x_{ti}^\top\bbeta)$ denotes a guiding loss term that characterizes the deviation of the incoming batch from the underlying response structure, where $\breve{y}_{ti}$ is a latent pseudo-response reflecting the response that would be generated under the true parameter. The first term captures the empirical information contained in the observed batch, whereas the second term, $\ploss_{n_t}(\bbeta)$, serves as a stabilizing anchor that regularizes the contribution of the incoming batch toward the underlying structural pattern. The adaptive weight $\gamma_t$ controls the trade-off between incorporating new information and maintaining consistency with the previously accumulated homogeneous batches.

Aggregating all batches up to $b$, the overall adaptive renewable loss function becomes:
\begin{equation}
\begin{aligned} 
\loss_{N_{b}}^a(\bbeta)=\frac{1}{N_b}\sum_{t=1}^bn_t\cdot\loss_{n_t}^a(\bbeta).
\end{aligned}
\label{eq:adp-loss}
\end{equation}
Similar to Eq.\eqref{eq:online-taylor}, Eq.\eqref{eq:adp-loss} can be approximated by
\begin{equation}\label{eq:adp-loss2}
\begin{aligned}
    \tilde{\mathcal{L}}_{N_b}^{a}(\boldsymbol{\beta})
    =
    \frac{1}{N_b}
    \Bigg\{
    & \frac{1}{2}
    \left(
    \boldsymbol{\beta}-\tilde{\boldsymbol{\beta}}_{b-1}^{a}
    \right)^\top\times
    \sum_{t=1}^{b-1}
    \left[
    \gamma_t n_t \nabla^2\mathcal{L}_{n_t}
    (\tilde{\boldsymbol{\beta}}_t^{a})
    +
    (1-\gamma_t)n_t
    \nabla^2\breve{\mathcal{L}}_{n_t}
    (\tilde{\boldsymbol{\beta}}_t^{a})
    \right] \\
    &\times
    \left(
    \boldsymbol{\beta}-\tilde{\boldsymbol{\beta}}_{b-1}^{a}
    \right)
    +n_b
    \left[
    \gamma_b \mathcal{L}_{n_b}(\boldsymbol{\beta})
    +
    (1-\gamma_b)\breve{\mathcal{L}}_{n_b}(\boldsymbol{\beta})
    \right]
    \Bigg\}.
\end{aligned}
\end{equation}
By differentiating \eqref{eq:adp-loss2} with respect to $\bbeta$ and replacing the unknown renewable estimator $\tilde{\boldsymbol{\beta}}_b^{a}$ in $\mathbf{W}_{n_b}(\tilde{\boldsymbol{\beta}}_b^{a})$, $\breve{\mathbf{W}}_{n_b}(\tilde{\boldsymbol{\beta}}_b^{a})$, $\mathbf{U}_{n_b}(\tilde{\boldsymbol{\beta}}_b^{a})$, and $\breve{\mathbf{U}}_{n_b}(\tilde{\boldsymbol{\beta}}_b^{a})$ with the previous-batch estimator $\tilde{\boldsymbol{\beta}}_{b-1}^{a}$, we obtain the following adaptive estimator, which accommodates potential cross-batch drift:
\begin{equation} \label{eq:areer}
\begin{aligned}
\tilde{\boldsymbol{\beta}}_b^{a}
=&
\Bigg[
\sum_{t=1}^{b-1}
\mathbf{W}_{n_t}^{a}(\tilde{\boldsymbol{\beta}}_t^{a})
+
\gamma_b \cdot\mathbf{W}_{n_b}(\tilde{\boldsymbol{\beta}}_{b-1}^{a})
+
(1-\gamma_b)\cdot\breve{\mathbf{W}}_{n_b}(\tilde{\boldsymbol{\beta}}_{b-1}^{a})
\Bigg]^{-1}
\\
&\times
\Bigg[
\sum_{t=1}^{b-1}
\mathbf{W}_{n_t}^{a}(\tilde{\boldsymbol{\beta}}_t^{a})
\tilde{\boldsymbol{\beta}}_{b-1}^{a}
+
\gamma_b \cdot\mathbf{U}_{n_b}(\tilde{\boldsymbol{\beta}}_{b-1}^{a})
+
(1-\gamma_b)\cdot\breve{\mathbf{U}}_{n_b}(\tilde{\boldsymbol{\beta}}_{b-1}^{a})
\Bigg],
\end{aligned}
\end{equation}
where $\mathbf{W}_{n_t}(\tilde{\boldsymbol{\beta}}^a)=\gamma_t \cdot\mathbf{W}_{n_t}(\tilde{\boldsymbol{\beta}}_{t}^{a})
+
(1-\gamma_t)\cdot\breve{\mathbf{W}}_{n_t}(\tilde{\boldsymbol{\beta}}_{t}^{a})$. 
More details on the derivation of the adaptive renewable estimator are provided in~\ref{areer-loss}.

The solution given in \eqref{eq:areer} adopts a different strategy from DReER. By introducing an adaptive weight, AReER automatically adjusts the contribution of newly arrived data. Specifically, the adaptive weight $\gamma_t$ shrinks the potential bias between the newly arrived data and the underlying true model, thereby downweighting the influence of potentially biased data batches. In addition, ARER preserves the computational advantages of the renewable framework, as the updating procedure relies only on the historical accumulated statistics $\left\{\sum_{t=1}^{b-1} \mathbf{W}_{n_t}^{a}(\tilde{\boldsymbol{\beta}}_t^{a}), \tilde{\boldsymbol{\beta}}_{b-1}^a\right\}$.

\begin{proposition}\label{P1}
Suppose that there exists a positive definite matrix $\mathbf A$ such that, uniformly over a neighborhood $\mathcal N(\boldsymbol{\beta}_0)$ of $\boldsymbol{\beta}_0$,
\begin{equation*}
    \sup_{\boldsymbol{\beta}\in\mathcal N(\boldsymbol{\beta}_0)}
    \left\|
    \frac1{n_t}\mathbf W_{n_t}(\boldsymbol{\beta})
    -
    \mathbf A
    \right\|
    =o_p(1),
    \qquad
    \sup_{\boldsymbol{\beta}\in\mathcal N(\boldsymbol{\beta}_0)}
    \left\|
    \frac1{n_t}\breve{\mathbf W}_{n_t}(\boldsymbol{\beta})
    -
    \mathbf A
    \right\|
    =o_p(1).
\end{equation*}
Under Regularity Conditions~\ref{con1}--\ref{con3}, suppose that the true parameter for batch $t$ is
$\boldsymbol{\beta}_t=\boldsymbol{\beta}_0+\boldsymbol{\eta}_t$,
where $\boldsymbol{\eta}_t$ denotes the batch-specific parameter deviation. In addition, suppose that the cumulative contribution of the historical estimation error is asymptotically negligible, that is,
\begin{equation*}
\frac1{N_b}
\sum_{t=1}^b
n_t(1-\gamma_t)
\mathbb E
\left(
\tilde{\boldsymbol{\beta}}_{t-1}^{a}
-
\boldsymbol{\beta}_0
\right)
=
o(1).
\end{equation*}
Then the AReER estimator satisfies
\begin{equation*}
\mathbb{E}
\left(
\tilde{\boldsymbol{\beta}}_b^{a}-\boldsymbol{\beta}_0
\right)
=
\frac1{N_b}
\sum_{t=1}^b
n_t\gamma_t\boldsymbol{\eta}_t
+
o(1).
\end{equation*}
Furthermore, define the weighted average batch-specific deviation as
$
\bar{\boldsymbol{\eta}}_b
=
\frac{
\sum_{t=1}^b n_t\gamma_t\boldsymbol{\eta}_t
}{
\sum_{t=1}^b n_t\gamma_t
}$.
Then
\begin{equation}\label{eq:pro1}
\mathbb{E}
\left(
\tilde{\boldsymbol{\beta}}_b^{a}-\boldsymbol{\beta}_0
\right)
=
\left(
\sum_{t=1}^b
\frac{n_t}{N_b}\gamma_t
\right)
\bar{\boldsymbol{\eta}}_b
+
o(1).
\end{equation}
\end{proposition}

Proposition \ref{P1} provides an intuitive characterization of the asymptotic bias of the AReER estimator under batchwise parameter heterogeneity. Specifically, the bias of $\tilde{\boldsymbol{\beta}}_b^a$ is asymptotically determined by a weighted average of the batch-specific parameter deviations $\boldsymbol{\eta}_t$. The weight $n_t\gamma_t/N_b$ reflects the effective contribution of batch $t$ to the final estimator: a larger batch size $n_t$ or a larger weighting factor $\gamma_t$ results in a greater influence of the corresponding parameter deviation on the estimator.

The result also highlights the role of $\gamma_t$ in controlling the influence of heterogeneous batches, which motivates the adaptive weighting strategy discussed later. When $\gamma_t$ is close to one, the deviation $\boldsymbol{\eta}_t$ of the current batch is largely incorporated into the estimator, thereby increasing its potential bias relative to the target parameter $\boldsymbol{\beta}_0$. In contrast, when $\gamma_t$ is close to zero, the direct contribution of the batch-specific deviation is substantially downweighted, thereby reducing its impact on the estimator. The assumption that the cumulative contribution of historical estimation errors is asymptotically negligible further implies that the leading-order bias is driven primarily by the weighted batch-specific deviations rather than by the propagation of historical estimation errors. 

Consequently, the asymptotic bias can be interpreted as shown in Eq.\eqref{eq:pro1}. Thus, AReER remains approximately unbiased when the weighted average deviation $\bar{\boldsymbol{\eta}}_b$ is close to zero, even when individual batches are heterogeneous. Conversely, a persistent nonzero weighted average deviation results in a non-negligible asymptotic bias. Additional details of the proof are provided in \ref{proof-sub3}. 

\subsection{Implementation of AReER}\label{sub:implementation-areer}

Finally, we discuss the construction of the pseudo-response and the adaptive weight $\gamma_t$. As with the AReER method, the true parameter is unavailable in practice and therefore cannot be directly used as the anchor for estimation. Following Theorem~1 of \cite{cao2026renewable}, the renewable estimator under the homogeneity assumption is consistent for the true parameter. Motivated by this result, we use the historical AReER estimator $\renew_{t-1}^a$ as a proxy for the true parameter to construct the pseudo-response:
$\breve{y}_{ti}=\x_{ti}^\top\renew_{t-1}^a$. Proposition~\ref{P1} further suggests that the adaptive weight should reflect the extent to which the incoming batch deviates from the historical data. The LM test statistic measures the discrepancy between the newly arrived batch and the underlying model represented by the historical estimator. Accordingly, we define the estimated adaptive weight as
\begin{equation*}
\hat{\gamma}_t
=
P\left\{
\chi_p^2
\geq
\widehat{\Lambda}_t\left(\renew_{t-1}^{a}\right)
\right\},
\end{equation*}
which corresponds to the upper-tail probability of the LM test statistic under the null hypothesis of homogeneity. Intuitively, $\hat{\gamma}_t$ measures the degree of compatibility between the incoming batch and the historical model. A smaller value of $\hat{\gamma}_t$ indicates stronger evidence of a batch-specific deviation, prompting the adaptive renewable estimator to place greater emphasis on the historical anchor and to downweight the contribution of the current batch.

When $\hat{\gamma}_t$ is close to zero, the current batch is assigned little weight in the standard update, and its contribution is dominated by the pseudo-response constructed from the historical anchor. Consequently, the updated estimator remains close to $\renew_{t-1}^{a}$. Conversely, as $\hat{\gamma}_t$ approaches one, the incoming batch is regarded as increasingly compatible with the historical model, and the resulting update approaches the standard renewable update. Given the estimated adaptive weight, the AReER estimator is defined as follows: 
\begin{equation} \label{eq:areer2}
\begin{aligned}
\tilde{\boldsymbol{\beta}}_b^{a}
=&
\Bigg[
\mathbf{H}_{N_{b-1}}^{a}(\tilde{\boldsymbol{\beta}}^{a})
+
\hat{\gamma}_b \cdot\mathbf{W}_{n_b}(\tilde{\boldsymbol{\beta}}_{b-1}^{a})
+
(1-\hat{\gamma}_b)\cdot\breve{\mathbf{W}}_{n_b}(\tilde{\boldsymbol{\beta}}_{b-1}^{a})
\Bigg]^{-1}
\\
&\times
\Bigg[
\mathbf{H}_{N_{b-1}}^{a}(\tilde{\boldsymbol{\beta}}^{a})
\tilde{\boldsymbol{\beta}}_{b-1}^{a}
+
\hat{\gamma}_b \cdot\mathbf{U}_{n_b}(\tilde{\boldsymbol{\beta}}_{b-1}^{a})
+
(1-\hat{\gamma}_b) \cdot \breve{\mathbf{U}}_{n_b}(\tilde{\boldsymbol{\beta}}_{b-1}^{a})
\Bigg].
\end{aligned}
\end{equation}
Here, $\mathbf{H}_{N_{b-1}}^{a}(\tilde{\boldsymbol{\beta}}^{a})=\sum_{t=1}^{b-1}
\mathbf{W}_{n_t}^{a}(\tilde{\boldsymbol{\beta}}_t^{a})$ denotes the adaptive cumulative Hessian matrix.

Compared with the hard-threshold monitoring procedure, the proposed adaptive scheme continuously accommodates potential deviations through adaptive weighting, thereby providing greater flexibility in handling moderate or subtle distributional shifts. The overall procedure of AReER is summarized in Algorithm~\ref{alg:areer}.

\begin{algorithm}[htbp]
\linespread{0.95}\selectfont
\caption{AReER: Adaptive Renewable Expectile Regression}
\label{alg:areer}
\SetKwInOut{Input}{Input}
\SetKwInOut{Output}{Output}

\Input{Streaming data $\mathcal{D}_{(b)}=\{\mathcal{D}_t\}_{t=1}^b$ with $\mathcal{D}_t=\{\boldsymbol{x}_{ti},y_{ti}\}_{i=1}^{n_t}$, expectile level $\tau$.}
\Output{Adaptive renewable expectile estimators under $\tau$ levle $\renew_t^a(\tau)$ and summary statistic $\mathcal{S}_t$ for $t=1,\dots,b$.}

\textbf{Initialization:}\\
$\hat{\boldsymbol{\beta}}_1$ using $\D_t$, and set $\hat{\boldsymbol{\beta}}_0=\hat{\boldsymbol{\beta}}_1$

Compute and store summary statistics:
\vspace{-12pt}
\[
\mathcal{S}_1=\{\mathbf{H}_{N_1}(\tilde{\boldsymbol{\beta}}^a), \tilde{\boldsymbol{\beta}}^a_1\}, \quad 
\mathbf{H}_{N_1}(\tilde{\boldsymbol{\beta}}^a)=\mathbf{W}_{n_1}(\tilde{\boldsymbol{\beta}}^a_1)=\sum_{i=1}^{n_1}|\tau-\mathbf{I}(y_{1i}<\boldsymbol{x}_{1i}^\top\tilde{\boldsymbol{\beta}}^a_1)|\boldsymbol{x}_{1i}\boldsymbol{x}_{1i}^\top.
\vspace{-12pt}
\]
\For{$t=2$ \KwTo $b$}{
    \textbf{Step 1: Compute adaptive weight $\gamma$.}\\
    Calculate the LM statistic $\widehat{\Lambda}_t(\bbeta)$ (Eq.\eqref{eq:LM-est}) and define
    $\hat{\gamma}_t = P(\chi^2_p \ge \widehat{\Lambda}_t(\bbeta_0))$.

    \textbf{Step 2: Adaptive renewable update.}\\
    Construct the pseudo-response 
    $$\breve{y}_{ti}=\x_{ti}^\top\renew_{t-1}.$$
    
    Update the adaptive renewable estimator $\renew_{t}^a$using Eq.\eqref{eq:areer2} and the pseudo-response $\breve{y}_{ti} = \x_{ti}^\top \renew_{t}^a.$

    Update cumulative statistics:
    \[ \mathbf{H}_{N_t}^a(\tilde{\boldsymbol{\beta}}^a)
    = \mathbf{H}_{N_{t-1}}^a(\tilde{\boldsymbol{\beta}}^a)
      + \hat{\gamma}_t\, \mathbf{W}_{n_t}(\tilde{\boldsymbol{\beta}}^a_t)
      + (1-\hat{\gamma}_t)\, \breve{\mathbf{W}}_{n_t}(\tilde{\boldsymbol{\beta}}^a_t).
    \]
    
    \textbf{Step 3: Store summary statistic $ \mathcal{S}_t=\left\{\mathbf{H}_{N_t}^a(\tilde{\boldsymbol{\beta}}^a), 
    \tilde{\boldsymbol{\beta}}^a_t\right\}$ and discard data batch $\D_t$.}\\

}
\end{algorithm}

\section{Robust Renewable Expectile Regression with Abnormal Batches}\label{sec:robust}

Note that ER is particularly sensitive to heavy-tailed distributions and outliers due to its quadratic-type loss function. This lack of robustness can be further exacerbated in streaming settings with abnormal batches, highlighting the need for a robust extension of the renewable estimator proposed in Section~\ref{sec:DReER} and Section~\ref{sec:AReER}. To address this issue, we extend the renewable estimation framework to a robust expectile regression setting.

Inspired by \citet{man2024retire}, we replace the quadratic component in the expectile loss with the Huber loss \citep{huber1964robust}, which provides a trade-off between $\ell_2$ and $\ell_1$ losses. Specifically, the Huber loss is
\begin{equation}
    \label{eq:huberloss}
    \rho^{\text{hub}}_\delta(u)=\frac{u^2}{2}\mathbf{I}(|u|\le \delta)+\left(\delta |u| - \frac{\delta^2}{2}\right)\mathbf{I}(|u|> \delta).
\end{equation}
The tuning parameter $\delta>0$ controls the transition between the quadratic and linear regimes. For small residuals ($|u|\le \delta$), the loss retains its quadratic form, thereby preserving statistical efficiency. For large residuals ($|u|>\delta$), the loss increases linearly rather than quadratically, thereby reducing the influence of outliers.

By incorporating the Huber loss in \eqref{eq:huberloss} into the asymmetric expectile framework, we define the batch-specific robust loss as follows:
\begin{equation} \label{eq:robust-loss}
\mathcal{L}^{\text{hub}}_{n_t}(\boldsymbol{\beta})
=
\frac{1}{n_t}\sum_{i=1}^{n_t}
\rho_{\tau,\delta_t}^{\text{hub}}(\epsilon_{ti})
=
\frac{1}{n_t}\sum_{i=1}^{n_t}
|\tau-\mathbf{I}(\epsilon_{ti}<0)|
\cdot
\rho^{\text{hub}}_{\delta_t}(\epsilon_{ti}),
\end{equation}
where $\delta_t$ denotes the batch-specific Huber threshold. Compared with the original quadratic expectile loss, Eq.\eqref{eq:robust-loss} preserves the asymmetric weighting structure induced by $\tau$ while down-weighting the influence of large residuals through the linear tail of the Huber loss. 

Let $\psi_{\tau}^{\mathrm{hub}}(u)=[\rho_{\tau}^{\text{hub}}(u)]'$ be the first-order derivative, i.e.,
\begin{equation*}
\psi_{\tau}^{\mathrm{hub}}(u)
=
w^{\mathrm{exp}}_\tau(u)\cdot
w^{\mathrm{hub}}_\delta(u)\cdot u,
\quad
w^{\mathrm{hub}}_\delta(u)
= 1 \cdot \mathbf{I}(|u|\leq\delta)+\frac{\delta}{|u|}\cdot \mathbf{I}(|u|>\delta),
\label{eq:huber-weight}
\end{equation*}
where $w^{\mathrm{exp}}_\tau(u)
=
|\tau-\mathbf{I}(u<0)|$ is the asymmetric expectile weight. Accordingly, the robust batch-specific gradient function is given by,
\begin{equation*}\label{eq:exphuber-score2}
\boldsymbol{g}_t^{\mathrm{hub}}(\boldsymbol{\beta})
=
-\frac{1}{n_t}
\sum_{i=1}^{n_t}
w^{\mathrm{exp}}_\tau(\epsilon_{ti})
w^{\mathrm{hub}}_{\delta_t}(\epsilon_{ti})
\epsilon_{ti}\boldsymbol{x}_{ti}.
\end{equation*}
Consequently, the detective-based renewable and adaptive procedures developed in the previous sections can be naturally extended to the robust renewable framework. 

Specifically, the corresponding local summary statistics are reformulated as:
\begin{equation}\label{eq:robust-stat}
\mathbf{W}_{n_t}^{\text{hub}}(\boldsymbol{\beta})
=
\sum_{i=1}^{n_t}
w^{\text{exp}}_\tau(\epsilon_{ti})
w^{\text{hub}}_{\delta_t}(\epsilon_{ti})
\boldsymbol{x}_{ti}\boldsymbol{x}_{ti}^{\top}, \quad
\mathbf{U}_{n_t}^{\text{hub}}(\boldsymbol{\beta})
=
\sum_{i=1}^{n_t}
w^{\text{exp}}_\tau(\epsilon_{ti})
w^{\text{hub}}_{\delta_t}(\epsilon_{ti})
\boldsymbol{x}_{ti}y_{ti}.
\end{equation}
The corresponding robust LM test statistic is defined as,
\begin{equation}\label{eq:robust-LM}
\Lambda_t^{\text{hub}}(\boldsymbol{\beta}_0)
=
n_t
\boldsymbol{g}_t^{\text{hub}}(\boldsymbol{\beta}_0)^{\top}
\left[
\widehat{\mathbf{C}}_t^{\text{hub}}(\boldsymbol{\beta}_0)
\right]^{-1}
\boldsymbol{g}_t^{\text{hub}}(\boldsymbol{\beta}_0).
\end{equation}
where $\epsilon_{ti}^0
=y_{ti}-\boldsymbol{x}_{ti}^{\top}\boldsymbol{\beta}_0$
denotes the residual under the null hypothesis $\boldsymbol{\beta}_0$. The corresponding covariance estimator of the score function is
$$
\widehat{\mathbf{C}}_t^{\text{hub}}(\boldsymbol{\beta}_0)
=
\frac{1}{n_t}
\sum_{i=1}^{n_t}
\left[
w^{\text{exp}}_\tau\left(\epsilon_{ti}^0\right)\cdot
w^{\text{hub}}_{\delta_t}\left(\epsilon_{ti}^0\right)\cdot
\epsilon_{ti}^0
\right]^2
\boldsymbol{x}_{ti}\boldsymbol{x}_{ti}^{\top}.
$$
With Eqs.\eqref{eq:robust-stat} and \eqref{eq:robust-LM} both the detective-based and adaptive weighting schemes can be readily extended to the robust renewable framework. Specifically, the robust LM statistic is first computed. For the monitoring-based procedure, the statistic is compared with the $\chi^2_{p,1-\alpha}$ critical value to determine whether an abnormal batch is detected. For the adaptive procedure, the corresponding right-tail probability is calculated and used as the adaptive weight $\gamma_t$. During estimation, the standard expectile summary statistics are replaced by their robust counterparts in Eq.\eqref{eq:robust-stat}, yielding the corresponding robust renewable estimator.

\begin{algorithm}[htbp]
    \linespread{0.95}\selectfont
    \caption{Robust Renewable Expectile Estimation}
    \label{alg:robust_reer}
    \SetKwInOut{Input}{Input}
    \SetKwInOut{Output}{Output}

    \Input{Streaming data $\mathcal{D}_{(b)}=\{\mathcal{D}_t\}_{t=1}^b$ with $\mathcal{D}_t=\{\boldsymbol{x}_{ti},y_{ti}\}_{i=1}^{n_t}$, expectile level $\tau$.}
    \Output{Robust renewable expectile estimators under $\tau$ level $\renew_t^\text{hub}(\tau)$ and summary statistic $\mathcal{S}_t$ for $t=1,\dots,b$.}
    
    \textbf{Initialization:}\\
    Compute the initial estimator $\boldsymbol{\beta}^{\text{(init)}}$ with OLS and obtained residuals:
    $
    \epsilon_{1i}^{\text{(init)}}.
    $\\
    Estimate the MAD-based residual scale  $\hat{\sigma}_1$, compute the Huber threshold $\delta_1 = 1.345\cdot \hat{\sigma}_1$.\\
    Compute robust estimator and store summary statistics:
    \begin{align*}
    &\tilde{\boldsymbol{\beta}}_{1}^{\mathrm{hub}}
    = \mathop{\arg\min}_{\boldsymbol{\beta}}
    \frac{1}{n_1}
    \sum_{i=1}^{n_1}
    \left|\tau-\mathbf{I}(\epsilon_{1i}<0)\right|
    \rho_{\delta_1}^{\mathrm{hub}}(\epsilon_{1i}),\\
    &\mathcal{S}_1
    =\left\{
    \mathbf{H}_{N_1}\left(\tilde{\boldsymbol{\beta}}_{1}^{\mathrm{hub}}\right),
    \tilde{\boldsymbol{\beta}}_{1}^{\mathrm{hub}},
    N_1=n_1
    \right\}, \quad
    \mathbf{H}_{N_1}\left(\tilde{\boldsymbol{\beta}}_{1}^{\mathrm{hub}}\right)
    =
    \mathbf{W}_{n_1}\!\left(\tilde{\boldsymbol{\beta}}_{1}^{\mathrm{hub}}\right).
    \end{align*}
    \For{$t=2$ \KwTo $b$}{
    \textbf{Step 1: Update the global scalar parameter}\\
    Calculate residuals with historical estimation $\tilde{\epsilon}_{ti} = y_{ti} - \boldsymbol{x}_{ti}^\top \tilde{\boldsymbol{\beta}}_{t-1}^{\text{hub}}$.\\
    Estimate residual scale $\hat{\sigma}_t$ and the adaptive threshold $\hat{\delta}_t = 1.345\cdot\hat{\sigma}_t$.

    \textbf{Step 2: Robust Renewable Estimation}\\
        According Eq.\eqref{eq:robust-LM} calculate the LM test statistic.\\
        Obtain renewable estimator with Algorithm~\ref{alg:dreer}~or Algorithm~\ref{alg:areer}.\\
        Update cummlative statistics:
        \[
        N_t = N_{t-1} + n_t,\quad
        \mathbf{H}_{N_t}(\tilde{\boldsymbol{\beta}}^{\text{hub}}) = \mathbf{H}_{N_{t-1}}(\tilde{\boldsymbol{\beta}}^{\text{hub}}) + \mathbf{W}_t^{\text{hub}}(\tilde{\boldsymbol{\beta}}_t^{\text{hub}}).
        \]

    \textbf{Step 3: Store summary statistic $\mathcal{S}_t = \{\mathbf{H}_{N_t}(\tilde{\boldsymbol{\beta}}^{\text{hub}}), \tilde{\boldsymbol{\beta}}_t^{\text{hub}}, N_t\}$ and discard data batch $\D_t$.}\\

}
\end{algorithm}

Finally, to ensure the scale adaptivity of the Huber threshold, we employ the median absolute deviation (MAD) to estimate the residual scale and automatically determine the threshold. For data batch $t\geq2$, we first compute the initial residuals $\tilde{\epsilon}_{ti}
=
y_{ti}
-
\boldsymbol{x}_{ti}^{\top}
\tilde{\boldsymbol{\beta}}_{t-1}^{\text{hub}}$ and then estimate the MAD-based residual scale as:
\begin{equation*}\label{eq:sigma-hat}
\hat{\sigma}_t
=
\frac{
\operatorname{median}_i
\left|
\tilde{\epsilon}_{ti}
-
\operatorname{median}
(\tilde{\epsilon}_{tj})
\right|
}{0.6745}.
\end{equation*}
Here, $0.6745=\Phi^{-1}(0.75)$ is the consistency constant under the standard normal assumption. Based on $\hat{\sigma}_t$, the Huber threshold is defined as $\hat{\delta}_t = k \hat{\sigma}_t$. The tuning parameter $k>0$ controls the transition between the $\ell_1$ and $\ell_2$ loss regimes. Throughout this paper, we set $k=1.345$, following the classical results in robust statistics. For $t=1$, since no historical estimator is available, we first obtain an initial estimate $\boldsymbol{\beta}^{(\text{init})}$ using ordinary least squares. The corresponding residuals are computed as $\epsilon_{1i}^{(\text{init})}
= y_{1i}
-\boldsymbol{x}_{1i}^{\top}
\boldsymbol{\beta}^{(\text{init})}$. Using the resulting threshold, the initial robust estimator $\tilde{\boldsymbol{\beta}}_1^{\text{hub}}$ is then obtained via IRLS algorithm. More details of the Robust Renewable Expectile Regression procedure are provided in Algorithm~\ref{alg:robust_reer}.

\section{Simulation Studies}\label{sec:experiments}

In this section, we conduct extensive simulation studies to evaluate the finite-sample performance of the proposed methods, including DReER and AReER and their robust versions, and compare them with several existing competitors. Specifically, we consider (a) ReER, the standard renewable expectile regression method proposed by \cite{cao2026renewable}; and (b) Oracle, an oracle version of ReER that has access to the true abnormal-batch information and serves as an ideal benchmark for assessing the performance of the proposed methods.

\subsection{Experimental Setup and Evaluation Metrics}\label{sub:4-1}

Suppose the samples $\{y_{ti},\x_{ti}\}$ are generated according to the model
\begin{align}\label{eq:model}
y_{ti} = \x_{ti}^{\top}\bbeta^{\star} + (\x_{ti}^{\top}\boldsymbol{\gamma})\e_{ti}, \quad i=1,\ldots,n_t; t = 1, \ldots, b,
\end{align}
where $\bbeta^{\star} = (1, 2, 1,1)^\top$. The covariate vector is specified as $\x_{ti}=(1,x_{ti,1},x_{ti,2},x_{ti,3})^\top$, where $[\x_{ti}]_{[1:3]}=[x_{ti,1},x_{ti,2},x_{ti,3}]^\top$ is independently generated from a truncated normal distribution $\mathcal{TN}(0,1;0,\infty)$. The parameter $\boldsymbol{\gamma}$ determines whether model \eqref{eq:model} has a homogeneous or heterogeneous error structure. Specifically, for the homogeneous scenario, we set $\boldsymbol{\gamma} = (1, 0, 0, 0)^\top$, whereas for the heterogeneous scenario, we set $\boldsymbol{\gamma} = (1, 0, 0.5, 0.5)^\top$. Under each setting, the random errors $\e_{ti}$ are generated from either a standard normal distribution, $\boldsymbol{\e} \sim \mathcal{N}(0,1)$, or a heavy-tailed Student's $t$ distribution with three degrees of freedom, $\boldsymbol{\e} \sim t(3)$. We consider four cases to represent different types of batch-level heterogeneity.

\begin{enumerate}[label=Case~\arabic*,leftmargin=*, itemindent=0pt, labelsep=0.5em,topsep=0pt, itemsep=0pt, parsep=0pt]
\item \label{simu:case1}\textbf{Random abnormal batches.}
The abnormal data are generated using an alternative parameter
$\bbeta^{\star}_{\mathrm{ab}}=\boldsymbol{\theta}
+u(1,1,1,1)^\top$,
where $\boldsymbol{\theta}=(0.4,-0.4,0.4,-0.4)^\top$ and
$u\sim\mathcal{U}(-0.01,0.01)$ controls the magnitude of the perturbation along the direction $(1,1,1,1)^\top$.
The index of an abnormal batch, denoted by $\iota$, is randomly selected from
$\{2,\ldots,b\}$. We consider two levels of abnormal-batch prevalence, with abnormal batches accounting for $10\%$ and $30\%$ of the total number of batches, respectively.
\item \label{simu:case2}\textbf{Block abnormal batches.}
The abnormal data are generated in the same manner as in \ref{simu:case1}.
The abnormal batches occur consecutively and occupy either the first 30\% or the last 30\% of all batches, with the proportion of abnormal batches fixed at 30\%.
\item \label{simu:case3}\textbf{Random abnormal batches with a local perturbation.}
The index of the abnormal batch, denoted by $\iota$, is randomly selected in the same manner as in \ref{simu:case1}, with the proportion of abnormal batches fixed at 30\%. Meanwhile, a local perturbation is introduced to the abnormal batch, $\bbeta^{\star}_{\mathrm{ab}}=\bbeta^\star+\boldsymbol{\theta}$ with
\(\boldsymbol{\theta}=\frac{2}{\sqrt{n_t}}(1,-1,1,-1)^\top\).
\item \label{simu:case4}\textbf{Abnormality with Gradual Drift.}
In this case, the true coefficient sequence follows a symmetric triangular drift pattern centered at $b_0=3b/4$. Specifically, the coefficient vector deviates linearly from the baseline parameter $\bbeta^\star$, reaches its maximum deviation $\boldsymbol{\theta}=(0.4,-0.4,0.4,-0.4)^\top$ at $b_0$, and then gradually returns to $\bbeta^\star$ at the end of the data stream. Accordingly, the true coefficient vector for the $t$-th batch is given by
$$
\bbeta_t = \bbeta^\star + \left( 1 - \frac{4\cdot\left|t - 3b/4\right|}{b} \right)_+ \boldsymbol{\theta}.
$$
Unlike the previous cases, where heterogeneity arises from abrupt changes across batches, the heterogeneity in this case evolves gradually over the data stream.
\end{enumerate}

For evaluation, each simulation setting was independently replicated 200 times. The performance was assessed using the mean squared error (MSE), defined as
\begin{equation*}
\text{MSE}(\beta_j)=\frac{1}{200}\sum_{s=1}^{200}\big(\tilde{\beta}^{(s)}_j-\beta_j\big)^2,
\end{equation*}
where $\tilde{\beta}_j^{(s)}$ denotes the renewable estimate of the $j$th regression coefficient obtained in the $s$th replication, and $\beta_{0j}(\tau)$ denotes the corresponding component of the true parameter vector
$\boldsymbol{\beta}_0(\tau)
=\boldsymbol{\beta}^{*}+\boldsymbol{\gamma}\,e_{\tau}(\epsilon)$.

\subsection{\ref{simu:case1}: Random Abnormal Batches}\label{sub:case1}

For \ref{simu:case1}, we consider two streaming-data scenarios. Scenario S1 examines the effect of the batch size $n_t$ on estimator performance by fixing the total sample size at $N_b=100{,}000$ and varying $n_t\in\{200,500,1000,2000\}$. Scenario S2 investigates the impact of the total number of batches $b$. Specifically, we set $n_1=300$ and fix the batch size at $n_t=200$ for all $t=2,3,\ldots,b$, while varying the total number of batches over $b\in\{50,100,200,500\}$. 

\subsubsection{Performance Estimation for Scenario 1}\label{case1-S1}

\begin{table}[htbp]
  \centering
  \renewcommand{\arraystretch}{0.8}
  \renewcommand\tabcolsep{9.0pt}
  \caption{Simulation results for \ref{simu:case1} at $\tau=0.25$, with fixed $N_b=100{,}000$, varying $n_t \in \{100, 200, 500, 1000\}$, and a fixed abnormal-batch proportion of 10\%.}
  \label{tab:case1-10}
  \small
  \begin{tabular}{cc|cccc|cccc}
  \toprule
  \multirow{3}[6]{*}{$\boldsymbol{\beta}$}
  & \multirow{3}[6]{*}{Method}
  & \multicolumn{8}{c}{$n_t$} \\
  \cmidrule{3-10}
  & & 200 & 500 & 1000 & 2000 & 200 & 500 & 1000 & 2000 \\
  \cmidrule{3-10}
  & & \multicolumn{4}{c|}{Homogeneous model, $\epsilon\sim N(0,1)$}
  & \multicolumn{4}{c}{Homogeneous model, $\epsilon\sim t(3)$} \\
  \midrule
  $\beta_0$ & ReER   & 0.161 & 0.168 & 0.165 & 0.173 & 0.187 & 0.201 & 0.191 & 0.200 \\
            & Oracle & 0.009 & \textbf{0.007} & \textbf{0.006} & \textbf{0.008} & 0.034 & \textbf{0.037} & \textbf{0.032} & 0.031 \\
            & DReER  & \textbf{\underline{0.008}} & \textbf{\underline{0.007}} & \textbf{\underline{0.006}} & \textbf{\underline{0.008}} & \textbf{\underline{0.032}} & \underline{0.038} & \textbf{\underline{0.032}} & \textbf{\underline{0.029}} \\
            & AReER  & 0.013 & 0.010 & 0.014 & 0.011 & 0.091 & 0.083 & 0.060 & 0.045 \\
  \addlinespace
  $\beta_1$ & ReER   & 0.168 & 0.168 & 0.166 & 0.173 & 0.181 & 0.170 & 0.182 & 0.178 \\
            & Oracle & 0.004 & 0.004 & \textbf{0.003} & 0.004 & \textbf{0.014} & \textbf{0.012} & 0.015 & \textbf{0.014} \\
            & DReER  & \textbf{\underline{0.003}} & \textbf{\underline{0.003}} & \textbf{\underline{0.003}} & \textbf{\underline{0.003}} & \underline{0.026} & \textbf{\underline{0.012}} & \textbf{\underline{0.014}} & \textbf{\underline{0.014}} \\
            & AReER  & 0.005 & 0.006 & 0.005 & 0.006 & 0.039 & 0.033 & 0.024 & 0.022 \\
  \addlinespace
  $\beta_2$ & ReER   & 0.145 & 0.141 & 0.145 & 0.141 & 0.164 & 0.151 & 0.162 & 0.159 \\
            & Oracle & \textbf{0.004} & \textbf{0.003} & 0.004 & \textbf{0.004} & \textbf{0.014} & \textbf{0.014} & 0.014 & 0.016 \\
            & DReER  & \textbf{\underline{0.004}} & \textbf{\underline{0.003}} & \textbf{\underline{0.003}} & \textbf{\underline{0.004}} & \underline{0.017} & \textbf{\underline{0.014}} & \textbf{\underline{0.013}} & \textbf{\underline{0.015}} \\
            & AReER  & 0.007 & 0.005 & 0.006 & 0.005 & 0.033 & 0.034 & 0.027 & 0.024 \\
  \addlinespace
  $\beta_3$ & ReER   & 0.167 & 0.169 & 0.173 & 0.169 & 0.177 & 0.182 & 0.178 & 0.188 \\
            & Oracle & \textbf{0.003} & \textbf{0.003} & \textbf{0.004} & \textbf{0.003} & \textbf{0.014} & \textbf{0.014} & 0.013 & 0.013 \\
            & DReER  & \textbf{\underline{0.003}} & \textbf{\underline{0.003}} & \textbf{\underline{0.004}} & \textbf{\underline{0.003}} & \underline{0.025} & \textbf{\underline{0.014}} & \textbf{\underline{0.012}} & \textbf{\underline{0.012}} \\
            & AReER  & 0.005 & 0.006 & 0.007 & 0.004 & 0.051 & 0.032 & 0.019 & 0.017 \\
  \midrule
  \multicolumn{2}{c|}{}
  & \multicolumn{4}{c|}{Heterogeneous model, $\epsilon\sim N(0,1)$}
  & \multicolumn{4}{c}{Heterogeneous model, $\epsilon\sim t(3)$} \\
  \midrule
  $\beta_0$ & ReER   & 0.164 & 0.178 & 0.169 & 0.183 & 0.247 & 0.268 & 0.239 & 0.252 \\
            & Oracle & 0.059 & \textbf{0.024} & \textbf{0.020} & \textbf{0.024} & \textbf{0.111} & \textbf{0.121} & \textbf{0.102} & 0.098 \\
            & DReER  & \textbf{\underline{0.051}} & \textbf{\underline{0.025}} & \textbf{\underline{0.021}} & \textbf{\underline{0.025}} & \textbf{\underline{0.177}} & \textbf{\underline{0.171}} & \textbf{\underline{0.114}} & \textbf{\underline{0.090}} \\
            & AReER  & 0.059 & 0.044 & 0.045 & 0.037 & 0.310 & 0.338 & 0.199 & 0.153 \\
  \addlinespace
  $\beta_1$ & ReER   & 0.168 & 0.170 & 0.167 & 0.177 & 0.208 & 0.187 & 0.205 & 0.201 \\
            & Oracle & \textbf{0.022} & \textbf{0.012} & 0.011 & 0.013 & \textbf{0.047} & \textbf{0.044} & \textbf{0.048} & 0.046 \\
            & DReER  & 0.034 & \textbf{\underline{0.012}} & \textbf{\underline{0.010}} & \textbf{\underline{0.012}} & \textbf{\underline{0.127}} & \textbf{\underline{0.067}} & \textbf{\underline{0.055}} & \textbf{\underline{0.044}} \\
            & AReER  & \textbf{\underline{0.022}} & 0.019 & 0.018 & 0.019 & 0.144 & 0.122 & 0.077 & 0.078 \\
  \addlinespace
  $\beta_2$ & ReER   & 0.167 & 0.158 & 0.169 & 0.163 & 0.226 & 0.203 & 0.230 & 0.218 \\
            & Oracle & 0.035 & \textbf{0.014} & \textbf{0.016} & \textbf{0.019} & \textbf{0.064} & 0.068 & 0.067 & 0.072 \\
            & DReER  & \textbf{\underline{0.019}} & \underline{0.016} & \textbf{\underline{0.016}} & \textbf{\underline{0.019}} & \textbf{\underline{0.070}} & \textbf{\underline{0.061}} & \textbf{\underline{0.059}} & \textbf{\underline{0.071}} \\
            & AReER  & 0.035 & 0.032 & 0.026 & 0.024 & 0.165 & 0.167 & 0.133 & 0.107 \\
  \addlinespace
  $\beta_3$ & ReER   & 0.162 & 0.169 & 0.178 & 0.169 & 0.216 & 0.221 & 0.209 & 0.223 \\
            & Oracle & 0.057 & \textbf{0.015} & \textbf{0.018} & \textbf{0.015} & \textbf{0.066} & \textbf{0.067} & \textbf{0.062} & \textbf{0.059} \\
            & DReER  & \textbf{\underline{0.054}} & \textbf{\underline{0.018}} & \textbf{\underline{0.019}} & \textbf{\underline{0.015}} & \textbf{\underline{0.257}} & \textbf{\underline{0.150}} & \textbf{\underline{0.077}} & \textbf{\underline{0.061}} \\
            & AReER  & 0.057 & 0.032 & 0.033 & 0.022 & 0.425 & 0.249 & 0.106 & 0.088 \\
  \bottomrule
  \end{tabular}%
  \begin{tablenotes}
    \footnotesize
    \item Metrics are reported in units of $10^{-2}$.
  \end{tablenotes}
\end{table}

Table~\ref{tab:case1-10} presents the simulation results for $\tau=0.25$ with the abnormal-batch proportion fixed at 10\%. For clarity, the best-performing method is highlighted in bold, while the best-performing method among the three renewable methods, namely ReER, DReER, and AReER, is underlined. The main findings are summarized as follows:
\begin{enumerate}[label=(\arabic*), leftmargin=*, topsep=0pt,itemsep=0pt,parsep=0pt]
     \item The ReER method consistently exhibits the poorest performance, with MSEs substantially larger, often by an order of magnitude, than those of the other approaches.
    \item Both DReER and AReER achieve substantial improvements over ReER, indicating that accounting for the influence of abnormal data batches is crucial for reliable estimation. Comparing the two strategies, DReER consistently produces results closer to those of the Oracle estimator. This suggests that directly filtering out abnormal data batches is more effective than downweighting them in mitigating the adverse effects of abnormal batches. 
\item Furthermore, the results are consistent across different error structures and error distributions, indicating the robustness of the proposed methods.
\end{enumerate}

\begin{figure}[H]
    \centering
    \includegraphics[width=.8\linewidth]{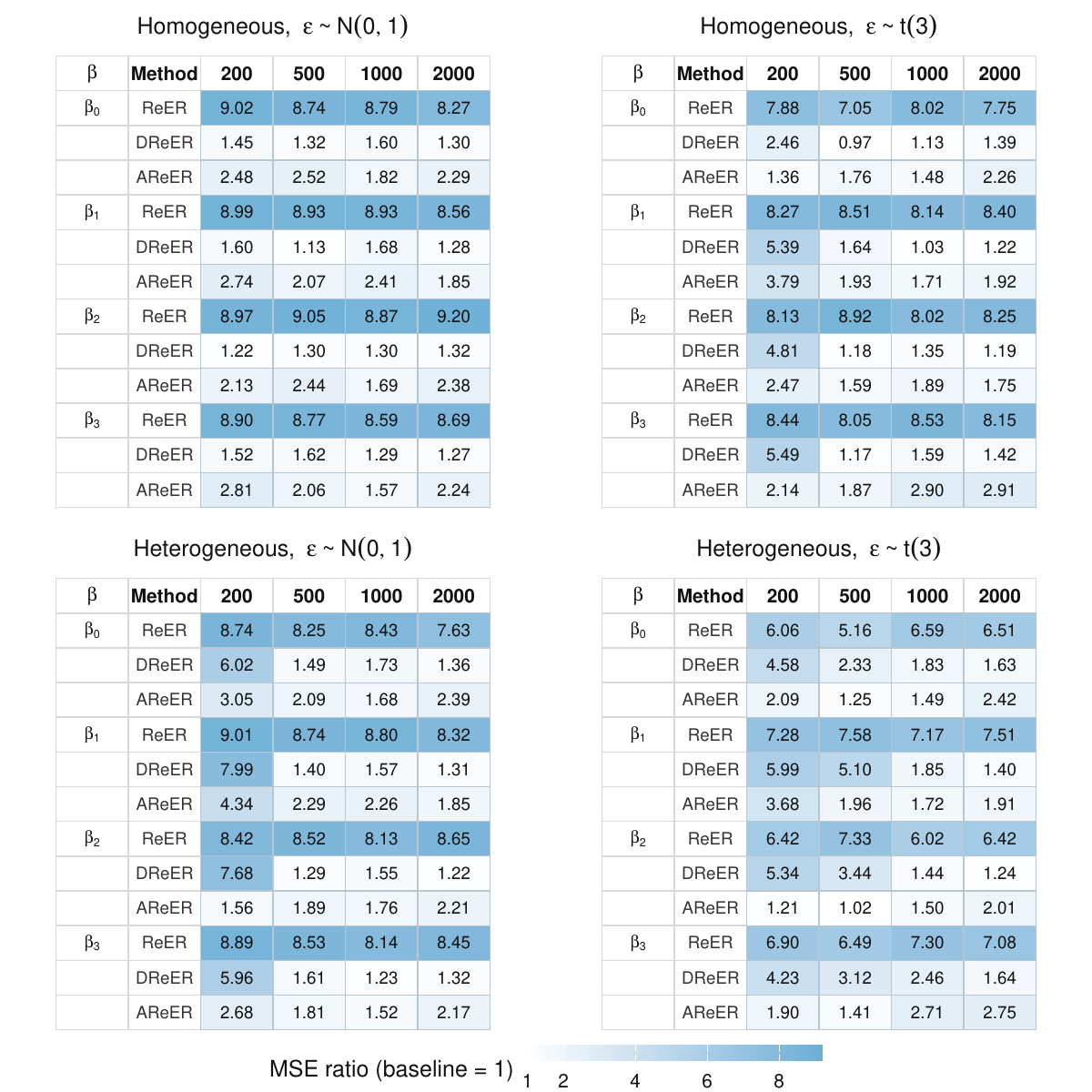}
    \caption{Ratio of MSEs (30\% abnormal proportion / 10\% abnormal proportion), with fixed $N_b=100{,}000$ and varying $n_t \in \{100, 200, 500, 1000\}$.}
    \label{fig:MSE_ratio}
    \centering
    \includegraphics[width=.75\linewidth]{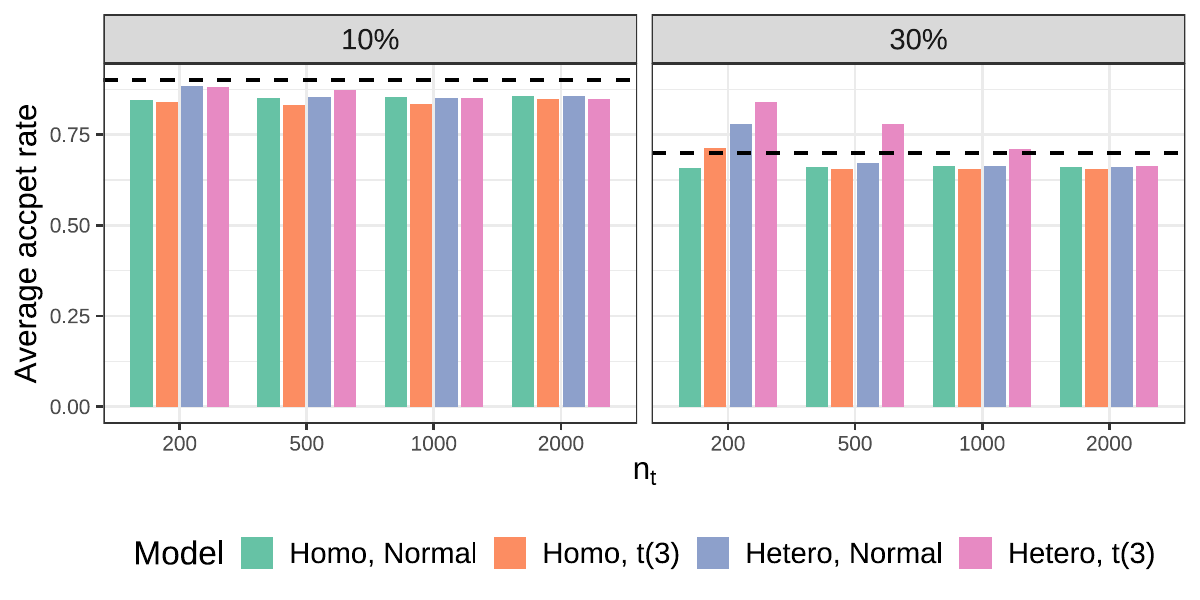}
    \caption{Proportion of effective samples used for estimation relative to the total sample size. The dotted black line denotes the true proportion of normal data.}
    \label{fig:n_rate}
\end{figure}

Figure~\ref{fig:MSE_ratio} reports the MSE ratios, defined as the MSE under a 30\% abnormal-batch proportion divided by that under a 10\% abnormal-batch proportion. The results show that, as the proportion of abnormal batches increases, the performance of ReER deteriorates substantially, resulting in a pronounced increase in MSE. This finding further highlights the importance of accounting for the potential influence of abnormal batches in online estimation. For DReER and AReER, we observe distinct performance patterns. Only under homogeneous model settings with normally distributed errors does DReER exhibit more stable performance than AReER, with the MSE ratios remaining close to one. This suggests that, when abnormal-batch detection is reliable, filtering out abnormal batches can be more effective than merely down-weighting their contributions. However, under heterogeneous model settings or with $t(3)$ errors, the performance of DReER deteriorates considerably when the batch size is small. In contrast, AReER does not exhibit such degradation, demonstrating greater robustness to increasing model complexity and decreasing batch sizes. 

Finally, Figure~\ref{fig:n_rate} displays the proportion of samples retained by DReER after the filtering procedure. We observe that, under simple model settings, the effective sample proportion closely matches the true proportion of normal samples. However, under scenarios with a high abnormal-batch proportion and small batch sizes $n_t$, the retained sample proportion exceeds the true proportion of normal samples when the data are generated under heterogeneous models or from heavy-tailed error distributions. This indicates that some abnormal batches are not successfully filtered out and consequently contribute to the online estimation, which helps explain the substantial increase in MSE observed for DReER.

\subsubsection{Performance Estimation for Scenario 2}\label{case1-S2}
\begin{figure}[H]
    \centering
    \includegraphics[width=.9\linewidth]{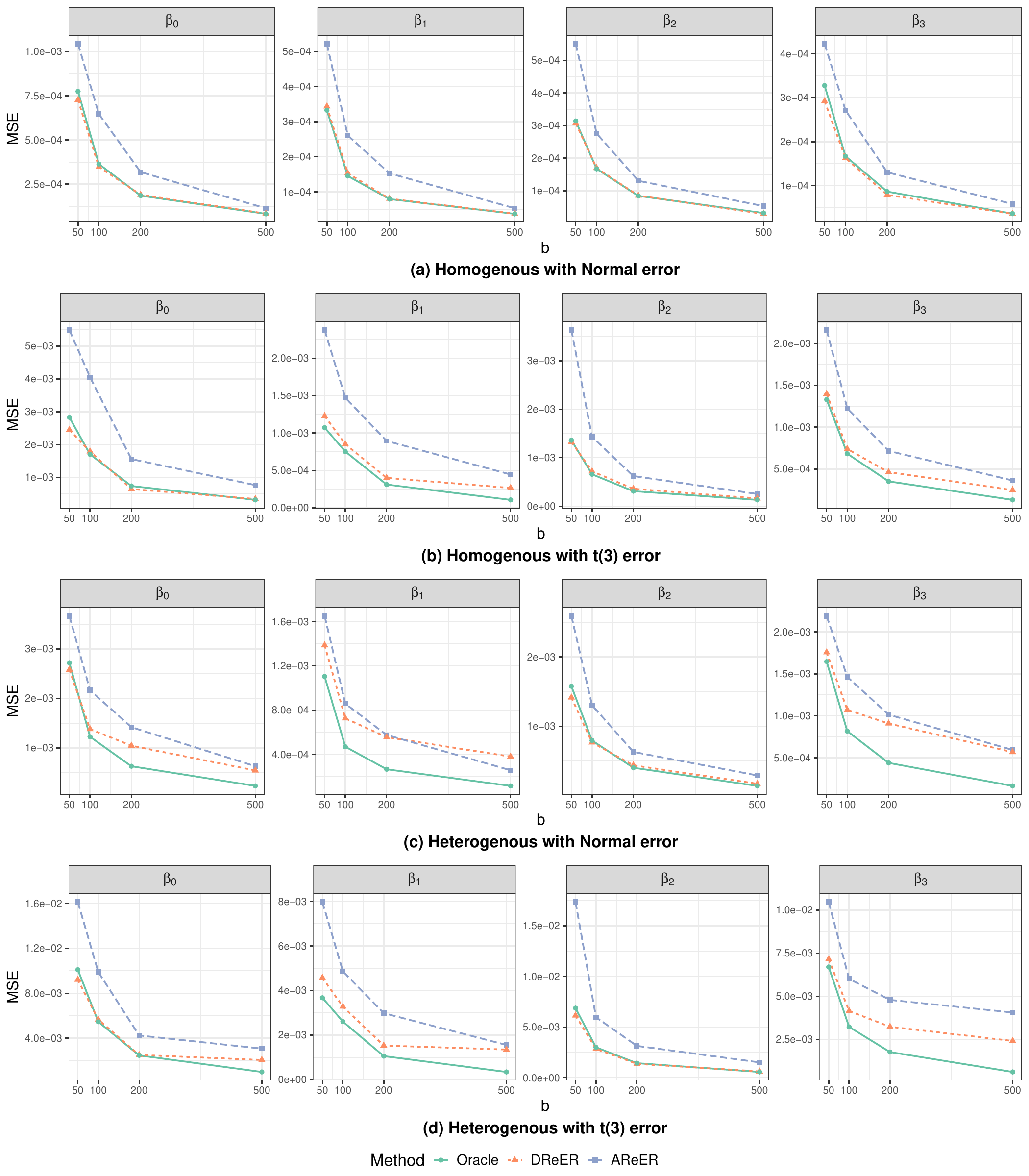}
    \caption{MSE values across different numbers of batches $b \in \{50,100,200,500\}$, with fixed batch size $n_t=200$ and an abnormal-batch proportion of 10\%.}
    \label{fig:MSE_K_10}
\end{figure}

For simplicity, we focus on the Oracle, DReER, and AReER methods and examine the performance of these three approaches that explicitly account for abnormal batches. Figure~\ref{fig:MSE_K_10}
shows that, as $b$ increases, the total sample size grows accordingly, leading to a gradual decrease in the MSE toward zero. This behavior provides empirical evidence supporting the consistency of these three methods. Moreover, in most cases, the performance of DReER is close to that of the Oracle benchmark, whereas AReER exhibits slightly inferior performance. This observation further indicates that, when the detection procedure is effective, the DReER method can achieve performance close to that of the Oracle benchmark.

\subsection{\ref{simu:case2}: Block Abnormal Batches}\label{sub:case2}
\ref{simu:case2} is designed to investigate whether the positions of abnormal batches in the data stream affect the performance of the proposed methods. The abnormal-batch proportion is fixed at 30\%. Table~\ref{tab:block30} and Table~\ref{tab:block70} report the MSEs when abnormal batches occur in the first 30\% and the last 70\% of the data stream, respectively. The results show that the MSEs are of comparable magnitude across the two settings, indicating that the positions of abnormal batches do not significantly affect the performance of DReER and AReER. In addition, consistent with the findings in \ref{simu:case1}, when the sample size is sufficiently large and the model is homogeneous, DReER outperforms AReER and achieves performance comparable to that of Oracle. Conversely, under heterogeneous model settings or heavy-tailed error distributions, AReER exhibits superior performance. Figures~\ref{fig:MSE_block_30} and \ref{fig:MSE_block_70} further confirm these observations.

\begin{figure}[htbp]
    \centering
    \begin{subfigure}{0.48\linewidth}
        
        \centering
        \includegraphics[width=\linewidth]{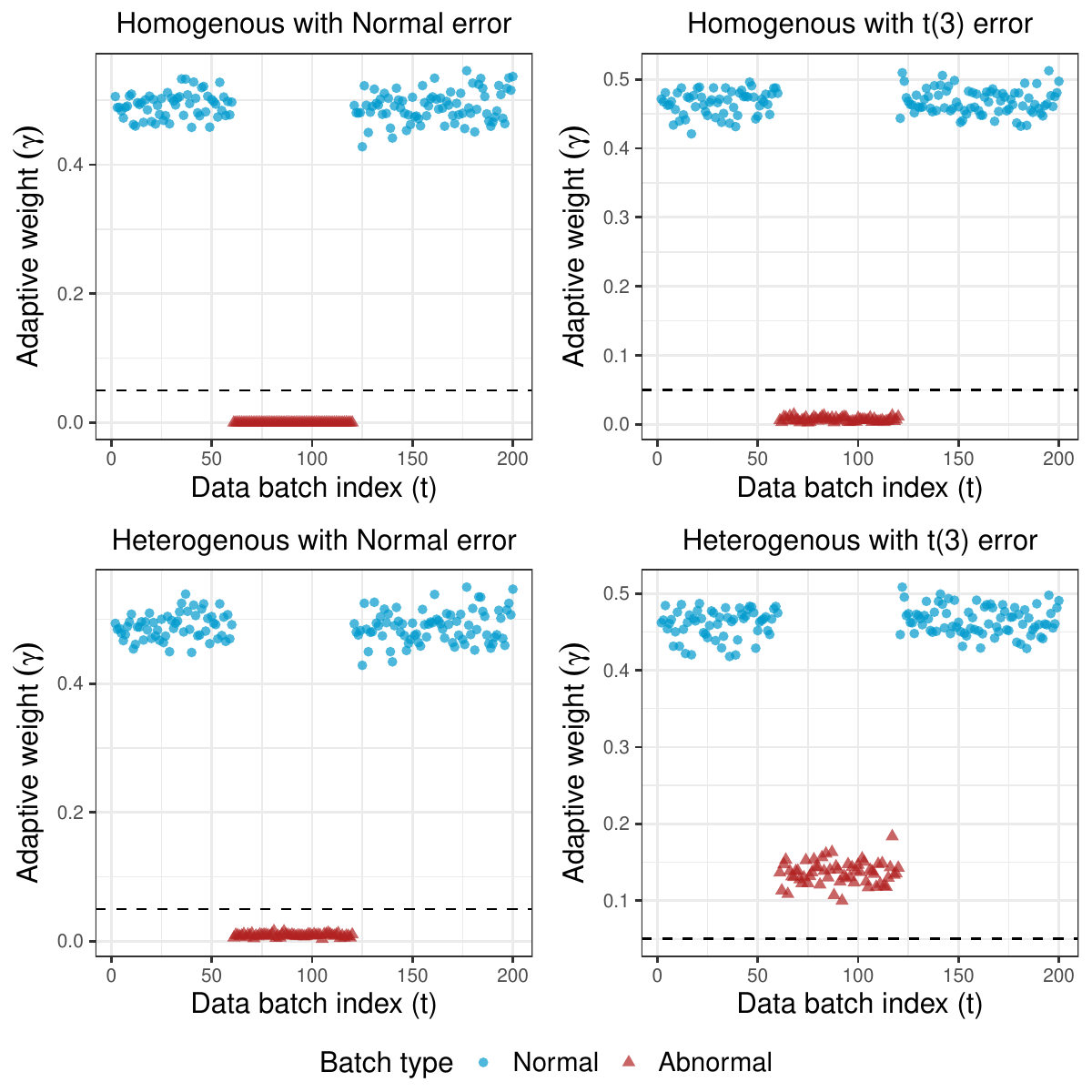}
        \caption{Abnormal batches in the initial 30\%.}
        \label{fig:gamma_block_30}
    \end{subfigure}
    \hfill
    \begin{subfigure}{0.48\linewidth}
        \centering
        \includegraphics[width=\linewidth]{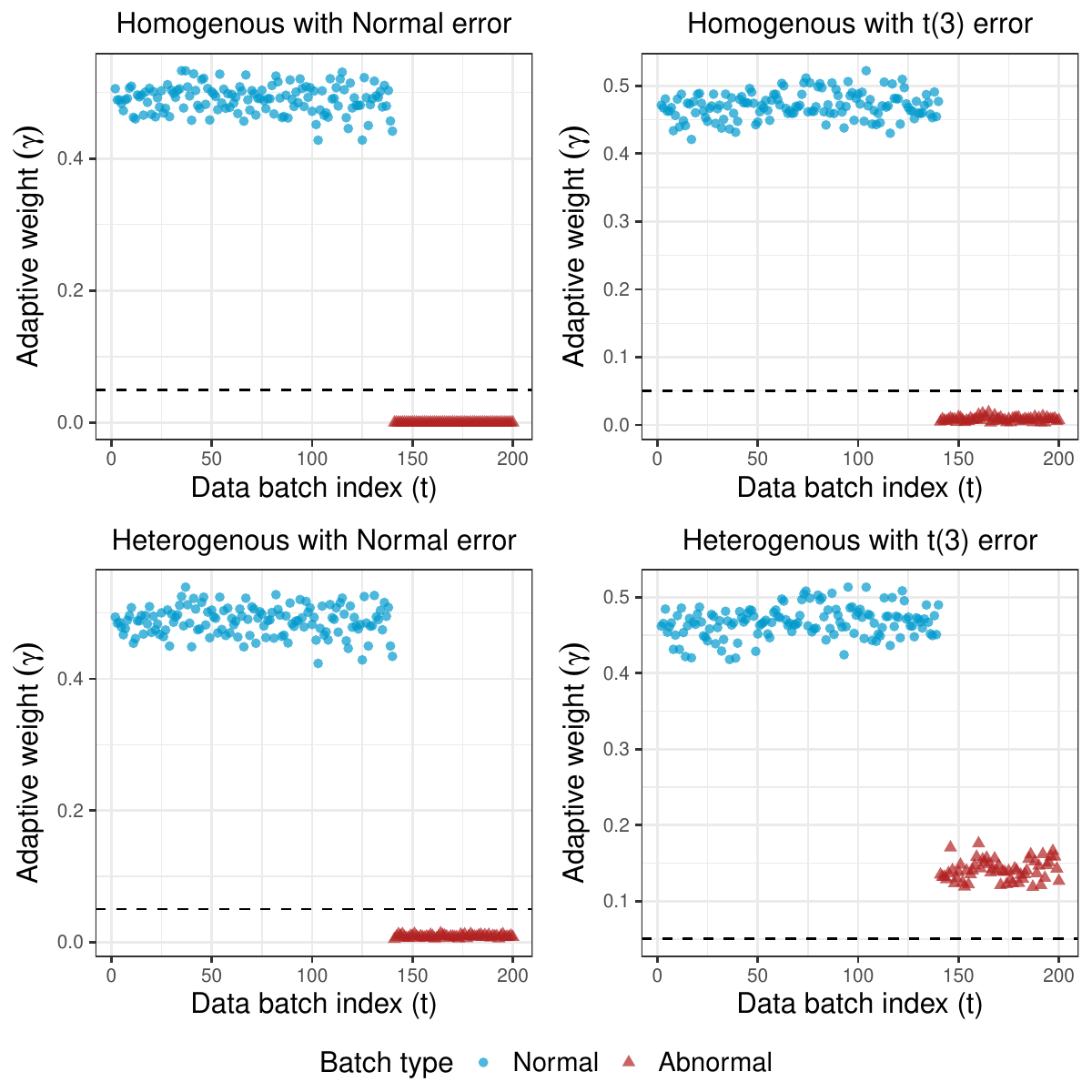}
       \caption{Abnormal batches in the last 70\%.}
        \label{fig:gamma_block_70}
    \end{subfigure}
    \caption{Detection signal/adaptive weight $\gamma_t$ under different locations of abnormal batches in the data stream.}
    \label{fig:gamma_block}
\end{figure}

To further investigate the underlying mechanism, we depict the detection signal or adaptive weight $\gamma_t$ for $n_t=500$ with fixed $N_b=100{,}000$, corresponding to $b=200$ batches. Figure~\ref{fig:gamma_block} shows that, under homogeneous model settings with normally distributed errors, the score statistic provides a clear separation between normal and abnormal batches, making the threshold-based detection procedure in DReER highly effective. However, when the model exhibits heterogeneity or the error distribution is heavy-tailed, the separation of the score statistic becomes substantially less distinct. In such cases, the performance of DReER becomes highly sensitive to the choice of the detection threshold. A fixed or pre-specified threshold may fail to adapt to changing distributional characteristics, resulting in either missed detections or excessive false alarms. This explains the observed degradation of DReER when the data-generating process deviates from the homogeneous Gaussian setting. In contrast, the adaptive weighting mechanism of AReER avoids hard classification decisions by continuously down-weighting potentially abnormal batches, resulting in more stable performance under heterogeneous models and heavy-tailed error distributions.

\subsection{\ref{simu:case3}: Random Abnormal Batches with a local perturbation }\label{sub:case3}

From the previous sections, we find that DReER achieves higher accuracy when the detection procedure is effective; however, it is also more prone to missed detections or excessive false alarms when the detection signal is weak. Motivated by this observation, we conduct simulations in which abnormal batches exhibit only local perturbations relative to normal batches, thereby further examining the robustness of the two methods under weak detection signals.

Tables~\ref{tab:MSE_small_nk} and \ref{tab:MSE_small_K} present the results under two streaming-data scenarios for \ref{simu:case3}. From these tables, we observe that, in most cases, AReER attains lower MSEs than DReER. In addition, Table~\ref{tab:MSE_small_nk} shows that the advantage of AReER gradually diminishes as the batch size $n_t$ increases. 

\begin{table}[htbp]\small
  \centering
\renewcommand{\arraystretch}{0.5}
  \setlength{\tabcolsep}{8pt}
  \caption{MSEs under \ref{simu:case3} with fixed $N_b=100{,}000$ and varying batch sizes $n_t \in \{100, 200, 500, 1000\}$.}
    \begin{tabular}{cc|cccc|cccc}
    \toprule
    \multirow{3}[6]{*}{$\bbeta$} & \multirow{3}[6]{*}{Method} & \multicolumn{4}{c|}{Homogeneous, $\e\sim \mathcal{N}(0,1)$} & \multicolumn{4}{c}{Homogeneous, $\e\sim t(3)$} \\
\cmidrule{3-10}          &       & \multicolumn{8}{c}{$n_t$} \\
\cmidrule{3-10}          &       & 200   & 500   & 1000  & 2000  & 200   & 500   & 1000  & 2000 \\
    \midrule
    \multirow{2}[1]{*}{$\beta_0$} & DReER   & 0.070  & 0.036  & 0.020  & \textbf{0.012}  & 0.079  & 0.044  & 0.050  & \textbf{0.044}  \\
          & AReR   & \textbf{0.032}  & \textbf{0.025}  & \textbf{0.018}  & 0.016  & \textbf{0.065}  & \textbf{0.058}  & \textbf{0.048}  & 0.053  \\
    \midrule
    \multirow{2}[0]{*}{$\beta_1$} & DReR   & 0.088  & 0.034  & 0.016  & 0.010  & 0.139  & 0.057  & 0.036  & 0.027  \\
          & AReER   & \textbf{0.035}  & \textbf{0.018}  & \textbf{0.010}  & \textbf{0.008}  & \textbf{0.117}  & \textbf{0.046}  & \textbf{0.026}  & \textbf{0.028}  \\
    \midrule
    \multirow{2}[0]{*}{$\beta_2$} & DReER   & 0.066  & 0.024  & 0.014  & \textbf{0.009}  & 0.098  & 0.051  & 0.028  & \textbf{0.021}  \\
          & AReER   & \textbf{0.022}  & \textbf{0.012}  & \textbf{0.007}  & \textbf{0.009}  & \textbf{0.051}  & \textbf{0.031}  & \textbf{0.026}  & 0.022  \\
    \midrule
    \multirow{2}[1]{*}{$\beta_3$} & DReER   & 0.083  & 0.033  & 0.016  & 0.009  & 0.139  & 0.057  & 0.044  & \textbf{0.028}  \\
          & AReER   & \textbf{0.031}  & \textbf{0.016}  & \textbf{0.010}  & \textbf{0.007}  & \textbf{0.094}  & \textbf{0.042}  & \textbf{0.037}  & 0.029  \\
    \midrule
    \multirow{3}[6]{*}{$\bbeta$} & \multirow{3}[6]{*}{Method} & \multicolumn{4}{c|}{Heterogeneous, $\e\sim \mathcal{N}(0,1)$} & \multicolumn{4}{c}{Heterogeneous, $\e\sim t(3)$} \\
\cmidrule{3-10}          &       & \multicolumn{8}{c}{$n_t$} \\
\cmidrule{3-10}          &       & 200   & 500   & 1000  & 2000  & 200   & 500   & 1000  & 2000 \\
    \midrule
    \multirow{2}[1]{*}{$\beta_0$} & DReER   & 0.187  & 0.097  & 0.059  & \textbf{0.034}  & \textbf{0.264}  & \textbf{0.139}  & \textbf{0.168}  & \textbf{0.135}  \\
          & AReER   & \textbf{0.161}  & \textbf{0.084}  & \textbf{0.048}  & 0.042  & 0.346  & 0.208  & 0.170  & 0.178  \\
    \midrule
    \multirow{2}[0]{*}{$\beta_1$} & DReER   & 0.162  & 0.067  & 0.036  & 0.023  & \textbf{0.214}  & \textbf{0.096}  & 0.071  & \textbf{0.063}  \\
          & AReER   & \textbf{0.094}  & \textbf{0.045}  & \textbf{0.026}  & \textbf{0.021}  & 0.245  & 0.112  & \textbf{0.070}  & 0.076  \\
    \midrule
    \multirow{2}[0]{*}{$\beta_2$} & DReER   & 0.090  & 0.043  & 0.029  & \textbf{0.023}  & \textbf{0.090}  & \textbf{0.072}  & \textbf{0.070}  & \textbf{0.061}  \\
          & AReER   & \textbf{0.029}  & \textbf{0.027}  & \textbf{0.022}  & 0.028  & 0.133  & 0.088  & 0.083  & 0.089  \\
    \midrule
    \multirow{2}[1]{*}{$\beta_3$} & DReER   & 0.194  & 0.085  & 0.049  & 0.028  & \textbf{0.360}  & \textbf{0.162}  & \textbf{0.136}  & \textbf{0.093}  \\
          & AReER   & \textbf{0.148}  & \textbf{0.066}  & \textbf{0.039}  & \textbf{0.024}  & 0.462  & 0.194  & 0.166  & 0.125  \\
    \bottomrule
    \end{tabular}%
    \captionsetup{justification=raggedright,singlelinecheck=false}
  \caption*{\footnotesize Metrics are reported in units of $10^{-2}$.}
  \label{tab:MSE_small_nk}%

  \centering
\renewcommand{\arraystretch}{0.5}
  \setlength{\tabcolsep}{8pt}
  \caption{MSEs under \ref{simu:case3} with fixed $n_t = 200$ and varying batch sizes $b \in \{50, 100, 200, 500\}$.}
    \begin{tabular}{cc|cccc|cccc}
    \toprule
    \multirow{3}[6]{*}{$\bbeta$} & \multirow{3}[6]{*}{Method} & \multicolumn{4}{c|}{Homogeneous, $\e\sim \mathcal{N}(0,1$)} & \multicolumn{4}{c}{Homogeneous, $\e\sim t(3)$} \\
\cmidrule{3-10}          &       & \multicolumn{8}{c}{$b$} \\
\cmidrule{3-10}          &       & 50    & 100   & 200   & 500   & 50    & 100   & 200   & 500 \\
    \midrule
    \multirow{2}[1]{*}{$\beta_0$} & DReER   & \textbf{0.124}  & 0.097  & 0.091  & 0.069  & \textbf{0.377}  & \textbf{0.210}  & \textbf{0.126}  & 0.082  \\
          & AReER   & 0.134  & \textbf{0.073}  & \textbf{0.061}  & \textbf{0.036}  & 0.474  & 0.272  & 0.145  & \textbf{0.060}  \\
    \midrule
    \multirow{2}[0]{*}{$\beta_1$} & DReR   & 0.112  & 0.100  & 0.087  & 0.081  & \textbf{0.259}  & \textbf{0.187}  & 0.144  & 0.141  \\
          & AReER   & \textbf{0.082}  & \textbf{0.046}  & 0.045  & 0.032  & 0.303  & 0.202  & \textbf{0.119}  & \textbf{0.091}  \\
    \midrule
    \multirow{2}[0]{*}{$\beta_2$} & DReER   & 0.091  & 0.086  & 0.068  & 0.066  & \textbf{0.173}  & \textbf{0.131}  & 0.123  & 0.092  \\
          & AReER   & \textbf{0.060}  & \textbf{0.051}  & \textbf{0.028} & \textbf{0.017}  & 0.286  & 0.168  & \textbf{0.095}  & \textbf{0.049}  \\
    \midrule
    \multirow{2}[1]{*}{$\beta_3$} & DReER   & 0.106  & 0.098  & 0.100  & 0.083  & \textbf{0.248}  & 0.171  & 0.158  & 0.131  \\
          & AReER   & \textbf{0.072}  & \textbf{0.053}  & \textbf{0.045}  & \textbf{0.033}  & 0.280  & \textbf{0.162}  & \textbf{0.123}  & \textbf{0.091}  \\
    \midrule
    \multirow{3}[6]{*}{$\bbeta$} & \multirow{3}[6]{*}{Method} & \multicolumn{4}{c|}{Heterogeneous, $\e\sim \mathcal{N}(0,1)$} & \multicolumn{4}{c}{Heterogeneous, $\e\sim t(3)$} \\
\cmidrule{3-10}          &       & \multicolumn{8}{c}{$b$} \\
\cmidrule{3-10}          &       & 50    & 100   & 200   & 500   & 50    & 100   & 200   & 500 \\
    \midrule
    \multirow{2}[1]{*}{$\beta_0$} & DReER   & \textbf{0.370}  & 0.276  & 0.237  & 0.181  & \textbf{1.299}  & \textbf{0.649}  & \textbf{0.412}  & \textbf{0.258}  \\
          & AReER   & 0.382  & \textbf{0.255}  & \textbf{0.227}  & \textbf{0.162}  & 1.649  & 0.915  & 0.589  & 0.353  \\
    \midrule
    \multirow{2}[0]{*}{$\beta_1$} & DReER   & 0.257  & 0.198  & 0.161  & 0.143  & \textbf{0.604} & \textbf{0.334} & \textbf{0.245}  & \textbf{0.220}  \\
          & AReER   & \textbf{0.218}  & \textbf{0.126}  & \textbf{0.116}  & \textbf{0.079}  & 0.843  & 0.527  & 0.283  & 0.193  \\
    \midrule
    \multirow{2}[0]{*}{$\beta_2$} & DReER   & 0.228  & 0.162  & 0.107  & 0.093  & \textbf{0.597}  & \textbf{0.294}  & \textbf{0.201}  & \textbf{0.083}  \\
          & AReER   & \textbf{0.214}  & \textbf{0.135}  & \textbf{0.064}  & \textbf{0.027}  & 1.318  & 0.628  & 0.326  & 0.158  \\
    \midrule
    \multirow{2}[1]{*}{$\beta_3$} & DReER   & 0.315  & 0.247  & 0.240  & 0.196  & \textbf{0.937}  & \textbf{0.524}  & \textbf{0.467}  & \textbf{0.335}  \\
          & AReER   & \textbf{0.289}  & \textbf{0.223}  & \textbf{0.192}  & \textbf{0.153}  & 1.394  & 0.744  & 0.705  & 0.473  \\
    \bottomrule
    \end{tabular}%
    \captionsetup{justification=raggedright,singlelinecheck=false}
  \caption*{\footnotesize Metrics are reported in units of $10^{-2}$.}
  \label{tab:MSE_small_K}%
\end{table}%

% \begin{figure}[htbp]
%     \centering
%     \includegraphics[width=1\linewidth]{figs/MSE_small_K.pdf}
%     \caption{Mean squared error (MSE) with fixed $n_t = 200$ and varying batch sizes $K \in \{50, 100, 200, 500\}$, under marginal deviations between normal and abnormal batches.}
%     \label{fig:MSE_small_K}
% \end{figure}

\subsection{\ref{simu:case4}: Abnormality with Gradual Drift}\label{sub:case4}

Finally, we present the simulation results for \ref{simu:case4} in Tables~\ref{tab:MSE_drift_nk} and \ref{tab:MSE_drift_K}. These results further indicate that, when abnormal batches do not exhibit sudden or substantial deviations from normal batches, detection-based methods may be prone to false detections due to weak statistical signals. In contrast, AReER provides more robust estimation by adaptively down-weighting information from potentially abnormal batches.

\begin{table}[htbp]\small
\centering
\renewcommand{\arraystretch}{0.5}
  \setlength{\tabcolsep}{8pt}
  \caption{MSEs under \ref{simu:case4} with fixed $N_b = 100{,}000$ and varying batch sizes $n_t \in \{100, 200, 500, 1000\}$.}
    \begin{tabular}{cc|cccc|cccc}
    \toprule
    \multirow{3}[6]{*}{$\bbeta$} & \multirow{3}[6]{*}{Method} & \multicolumn{4}{c|}{Homogeneous, $\e\sim \mathcal{N}(0,1$)} & \multicolumn{4}{c}{Homogeneous, $\e\sim t(3)$} \\
\cmidrule{3-10}          &       & \multicolumn{8}{c}{$n_t$} \\
\cmidrule{3-10}          &       & 200   & 500   & 1000  & 2000  & 200   & 500   & 1000  & 2000 \\
    \midrule
    \multirow{2}[1]{*}{$\beta_0$} & DReER   & 0.080  & 0.021  & \textbf{0.013}  & \textbf{0.013}  & 0.246  & 0.125  & \textbf{0.064}  & \textbf{0.053}  \\
          & AReER   & \textbf{0.018}  & \textbf{0.014}  & 0.016  & 0.015  & \textbf{0.065}  & \textbf{0.083}  & \textbf{0.064}  & 0.086  \\
    \midrule
    \multirow{2}[0]{*}{$\beta_1$} & DReER   & 0.092  & 0.018  & 0.010  & \textbf{0.006}  & 0.369  & 0.134  & 0.054  & \textbf{0.029}  \\
          & AReER   & \textbf{0.016}  & \textbf{0.007}  & \textbf{0.007}  & \textbf{0.006}  & \textbf{0.082}  & \textbf{0.044}  & \textbf{0.028}  & 0.036  \\
    \midrule
    \multirow{2}[0]{*}{$\beta_2$} & DReER   & 0.071  & 0.016  & 0.009  & \textbf{0.006}  & 0.247  & 0.099  & 0.047  & \textbf{0.033}  \\
          & AReER   & \textbf{0.009}  & \textbf{0.007}  & \textbf{0.007}  & 0.008  & \textbf{0.045}  & \textbf{0.035}  & \textbf{0.036}  & 0.035  \\
    \midrule
    \multirow{2}[1]{*}{$\beta_3$} & DReER   & 0.092  & 0.020  & 0.009  & \textbf{0.007}  & 0.355  & 0.149  & 0.053  & 0.038  \\
          & AReER   & \textbf{0.019} & \textbf{0.007}  & \textbf{0.007}  & 0.008  & \textbf{0.085}  & \textbf{0.051}  & \textbf{0.032}  & \textbf{0.033}  \\
    \midrule
    \multirow{3}[6]{*}{$\bbeta$} & \multirow{3}[6]{*}{Method} & \multicolumn{4}{c|}{Heterogeneous, $\e\sim \mathcal{N}(0,1)$} & \multicolumn{4}{c}{Heterogeneous, $\e\sim t(3)$} \\
\cmidrule{3-10}          &       & \multicolumn{8}{c}{$n_t$} \\
\cmidrule{3-10}          &       & 200   & 500   & 1000  & 2000  & 200   & 500   & 1000  & 2000 \\
    \midrule
    \multirow{2}[1]{*}{$\beta_0$} & DReER   & 0.546  & 0.193  & 0.076  & 0.052  & 0.860  & 0.680  & 0.440  & 0.270  \\
          & AReER   & \textbf{0.161}  & \textbf{0.064}  & \textbf{0.049}  & \textbf{0.049}  & \textbf{0.458}  & \textbf{0.403}  & \textbf{0.222}  & \textbf{0.264}  \\
    \midrule
    \multirow{2}[0]{*}{$\beta_1$} & DReER   & 0.468  & 0.154  & 0.063  & 0.032  & 0.768  & 0.522  & 0.339  & 0.186  \\
          & AReER   & \textbf{0.090}  & \textbf{0.031}  & \textbf{0.026}  & \textbf{0.020}  & \textbf{0.240}  & \textbf{0.171}  & \textbf{0.103}  & \textbf{0.121}  \\
    \midrule
    \multirow{2}[0]{*}{$\beta_2$} & DReER   & 0.317  & 0.122  & 0.059  & 0.035  & 0.352  & 0.314  & 0.256  & 0.172  \\
          & AReER   & \textbf{0.031}  & \textbf{0.029}  & \textbf{0.029}  & \textbf{0.034}  & \textbf{0.161}  & \textbf{0.114}  & \textbf{0.140}  & \textbf{0.143}  \\
    \midrule
    \multirow{2}[1]{*}{$\beta_3$} & DReER   & 0.576  & 0.201  & 0.070  & 0.044  & 1.035  & 0.743  & 0.435  & 0.248  \\
          & AReER   & \textbf{0.172}  & \textbf{0.056} & \textbf{0.033}  & \textbf{0.033}  & \textbf{0.572}  & \textbf{0.360}  & \textbf{0.181}  & \textbf{0.161} \\
    \bottomrule
    \end{tabular}%
    \captionsetup{justification=raggedright,singlelinecheck=false}
  \caption*{\footnotesize Metrics are reported in units of $10^{-2}$.}
  \label{tab:MSE_drift_nk}%
\end{table}%

\begin{table}[htbp]\small
  \centering
\renewcommand{\arraystretch}{0.5}
  \setlength{\tabcolsep}{8pt}
  \caption{MSEs under \ref{simu:case4} with fixed $n_t = 200$ and varying batch sizes $b \in \{50, 100, 200, 500\}$.}
    \begin{tabular}{cc|cccc|cccc}
    \toprule
    \multirow{3}[6]{*}{$\bbeta$} & \multirow{3}[6]{*}{Method} & \multicolumn{4}{c|}{Homogeneous, $\e\sim \mathcal{N}(0,1)$} & \multicolumn{4}{c|}{Homogeneous, $\e\sim t(3)$} \\
\cmidrule{3-10}          &       & \multicolumn{8}{c}{$b$} \\
\cmidrule{3-10}          &       & 50   & 100   & 200  & 500  & 50   & 100   & 200  & 500 \\
    \midrule
    \multirow{2}[1]{*}{$\beta_0$} & DReER   & \textbf{0.168}  & 0.120  & 0.091  & 0.081  & \textbf{0.444}  & 0.296  & 0.236  & 0.208  \\
          & AReER   & \textbf{0.168}  & \textbf{0.087}  & \textbf{0.046}  & \textbf{0.021}  & 0.578  & \textbf{0.293}  & \textbf{0.140}  & \textbf{0.067}  \\
    \midrule
    \multirow{2}[0]{*}{$\beta_1$} & DReER  & 0.143  & 0.121  & 0.099  & 0.095  & 0.460  & 0.374  & 0.342  & 0.337  \\
          & AReER   & \textbf{0.085}  & \textbf{0.052}  & \textbf{0.033}  & \textbf{0.021}  & \textbf{0.337}  & \textbf{0.207}  & \textbf{0.126}  & \textbf{0.081}  \\
    \midrule
    \multirow{2}[0]{*}{$\beta_2$} & DReER   & 0.118  & 0.093  & 0.078  & 0.070  & 0.411  & 0.302  & 0.275  & 0.264  \\
          & AReER   & \textbf{0.081}  & \textbf{0.043}  & \textbf{0.023}  & \textbf{0.011}  & \textbf{0.385}  & \textbf{0.193}  & \textbf{0.104}  & \textbf{0.051}  \\
    \midrule
    \multirow{2}[1]{*}{$\beta_3$} & DReER   & 0.124  & 0.107  & 0.094  & 0.089  & 0.458  & 0.355  & 0.335  & 0.333  \\
          & AReER   & \textbf{0.071}  & \textbf{0.038}  & \textbf{0.025} & \textbf{0.016}  & \textbf{0.382} & \textbf{0.218}  & \textbf{0.137}  & \textbf{0.088}  \\
    \midrule
    \multirow{3}[6]{*}{$\bbeta$} & \multirow{3}[6]{*}{Method} & \multicolumn{4}{c|}{Heterogeneous, $\e\sim \mathcal{N}(0,1)$} & \multicolumn{4}{c}{Heterogeneous, $\e\sim t(3)$} \\
\cmidrule{3-10}          &       & \multicolumn{8}{c}{$b$} \\
\cmidrule{3-10}          &       & 50   & 100   & 200  & 500  & 50   & 100   & 200  & 500 \\
    \midrule
    \multirow{2}[1]{*}{$\beta_0$} & DReER   & 0.790  & 0.646  & 0.576  & 0.547  & \textbf{1.714}  & \textbf{1.095}  & 0.868  & 0.733  \\
          & AReER   & \textbf{0.574}  & \textbf{0.348}  & \textbf{0.243}  & \textbf{0.178}  & 2.238  & 1.157  & \textbf{0.695}  & \textbf{0.442}  \\
    \midrule
    \multirow{2}[0]{*}{$\beta_1$} & DReER   & 0.669  & 0.561  & 0.514  & 0.483  & 1.118  & 0.874  & 0.746  & 0.682  \\
          & AReER   & \textbf{0.314}  & \textbf{0.200}  & \textbf{0.146}  & \textbf{0.108}  & \textbf{1.034}  & \textbf{0.616}  & \textbf{0.381}  & \textbf{0.239}  \\
    \midrule
    \multirow{2}[0]{*}{$\beta_2$} & DReER   & 0.525  & 0.403  & 0.356  & 0.313  & \textbf{0.941}  & \textbf{0.591}  & \textbf{0.457}  & 0.367  \\
          & AReER   & \textbf{0.285}  & \textbf{0.155}  & \textbf{0.079}  & \textbf{0.036}  & 1.808  & 0.883  & 0.466  & \textbf{0.192}  \\
    \midrule
    \multirow{2}[1]{*}{$\beta_3$} & DReER   & 0.694  & 0.626  & 0.593  & 0.566  & \textbf{1.560}  & \textbf{1.196}  & 1.046  & 0.977  \\
          & AReER   & \textbf{0.352}  & \textbf{0.242}  & \textbf{0.188}  & \textbf{0.161}  & 2.025  & 1.243  & \textbf{0.859}  & \textbf{0.628}  \\
    \bottomrule
    \end{tabular}%
    \captionsetup{justification=raggedright,singlelinecheck=false}
  \caption*{\footnotesize Metrics are reported in units of $10^{-2}$.}
  \label{tab:MSE_drift_K}%
\end{table}%

% \begin{figure}[htbp]
%     \centering
%     \includegraphics[width=1\linewidth]{figs/MSE_drift_K.pdf}
%     \caption{MSEs with fixed $n_t = 200$ and varying batch sizes $K \in \{50, 100, 200, 500\}$, under gradual drift.}
%     \label{fig:MSE_drift_K}
% \end{figure}

\subsection{Simulation Results for Robust Renewable Expectile Regression}\label{sub:robust}

%In the previous section, we evaluated the proposed methods under heterogeneous streaming data. 

Although the results in Sections~\ref{sub:case1}–\ref{sub:case4} show that both DReER and AReER outperform ReER by effectively mitigating the impact of abnormal batches, their performance deteriorates under heavy-tailed error distributions (e.g., $t(3)$). This deterioration reflects the sensitivity of expectile regression to heavy-tailed errors and extreme observations. Although increasing the batch size $n_t$ can alleviate this issue, batch sizes are typically fixed and limited in practice. Therefore, we further examine two robust variants based on robust expectile regression to improve performance under heavy-tailed error distributions and in the presence of outliers. 

\begin{figure}[htbp]
    \centering
    \includegraphics[width=1\linewidth]{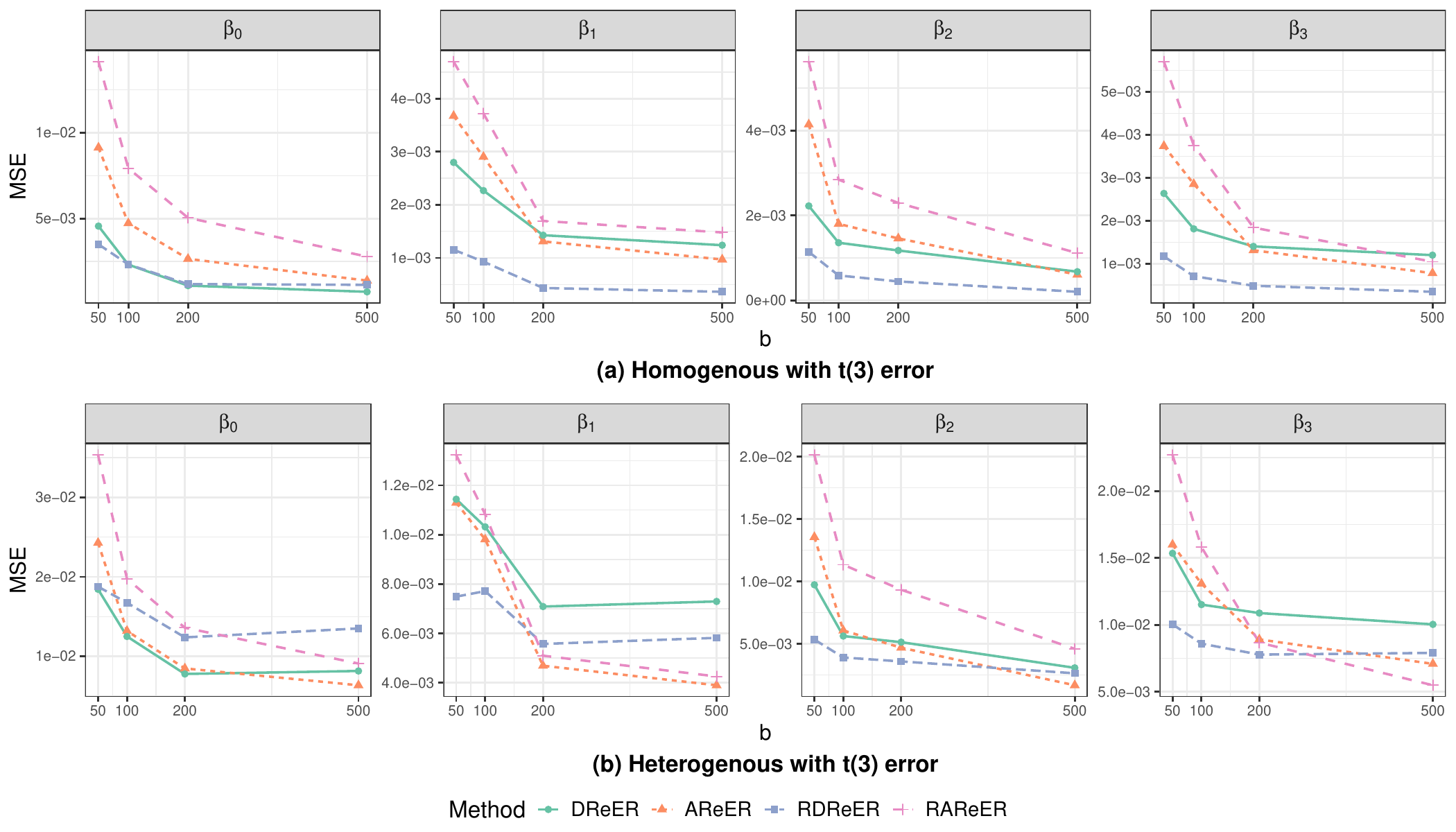}
    \caption{MSEs of robust renewable methods with fixed batch size $n_t=200$ and varying numbers of batches $b \in {50,100,200,500}$ under \ref{simu:case1}, with an abnormal-batch proportion of 30\%.}
    \label{fig:MSE_big_R}
\end{figure}

Figure~\ref{fig:MSE_big_R} presents the MSE values for \ref{simu:case1} with an abnormal-batch proportion of 30\%. Compared with the standard ER-based methods, the robust ER renewable approaches, namely RDReER and RAReER, achieve lower MSEs for most coefficients, except for $\beta_0$. This finding suggests that replacing the asymmetric least-squares loss in ER with a robust loss function, such as the Huber loss, can effectively mitigate the influence of outliers. Similar patterns are observed in Cases~\ref{simu:case3} and \ref{simu:case4}, which involve marginal heterogeneity and gradual drift, respectively, where the robust renewable methods consistently yield lower MSEs. Results for the remaining cases are reported in Figures~\ref{fig:MSE_small_R} and~\ref{fig:MSE_drift_R}.

Finally, we consider a more complex scenario in which the response $\boldsymbol{y}_t$ in batch $t$ is contaminated by random outliers:
\begin{align}\label{eq:modelout}
y_{ti}^o
= \x_{ti}^{\top}\bbeta^{\star}
+ (\x_{ti}^{\top}\boldsymbol{\gamma})\varepsilon_{ti}
+ \Delta_{ti}, \quad t = 1, \ldots, b,
\end{align}
where $\Delta_{ti} = \delta_{ti} \times 10 \cdot \mathrm{sd}(\boldsymbol{y}_t),
\delta_{ti}\in\{-1,1\}$ is generated with equal probability, and $\mathrm{sd}(\boldsymbol{y}_t)$ denotes the standard deviation of the response generated from the uncontaminated model. The proportion of contaminated responses is set to 5\%, and the abnormal-batch settings are consistent with those in \ref{simu:case1}.
The results are summarized in Table~\ref{tab:robust_out}. It can be seen that the standard ER-based method is substantially affected by the presence of outliers, particularly for the intercept, which is more directly influenced by $\Delta_{ti}$. In contrast, the robust ER methods effectively mitigate this issue, achieving MSEs below $0.01$ for all coefficients. These results demonstrate the robustness of the proposed robust ER approaches in the presence of outliers. 

\subsection{Key Findings}\label{sec:keyfinding}

For clarity, we summarize the key findings as follows:
\begin{itemize}[leftmargin=*, itemindent=0pt, labelsep=0.5em,topsep=0pt, itemsep=0pt, parsep=0pt]
    \item[1)] When heterogeneous batches are present, properly accounting for their impact substantially improves estimation accuracy. This improvement remains consistent across different heterogeneous model settings. Moreover, the positions of abnormal batches do not significantly affect the performance of either the detection-based (DReER) method or the adaptive (AReER) method, as demonstrated in Section~\ref{sub:case2}.
    \begin{table}[H]\small
  \centering
\renewcommand{\arraystretch}{0.65}
  \setlength{\tabcolsep}{8pt}
  \caption{MSEs with fixed batch size $n_t=200$ and varying numbers of batches $b \in \{50,100,200,500\}$, with data generated according to model\eqref{eq:modelout}.}
    \begin{tabular}{cc|cccc|cccc}
    \toprule
    \multirow{3}[6]{*}{$\bbeta$} & \multirow{3}[6]{*}{Method} & \multicolumn{4}{c|}{Homogeneous, $\e\sim \mathcal{N}(0,1)$} & \multicolumn{4}{c}{Homogeneous, $\e\sim t(3)$} \\
\cmidrule{3-10}          &       & \multicolumn{8}{c}{$b$} \\
\cmidrule{3-10}          &       & 50    & 100   & 200   & 500   & 50    & 100   & 200   & 500 \\
    \midrule
    \multirow{4}[1]{*}{$\beta_0$} & DReER   & 0.237  & 0.224  & 0.214  & 0.218  & 0.494  & 0.418  & 0.423  & 0.397  \\
          & RDReER  & \textbf{0.007}  & \textbf{0.006}  & \textbf{0.006}  & 0.006  & \textbf{0.002}  & \textbf{0.001}  & \textbf{0.001}  & \textbf{0.000}  \\
          & AReER   & 0.242  & 0.189  & 0.174  & 0.167  & 0.472  & 0.320  & 0.317  & 0.294  \\
          & RAReER  & 0.010  & 0.007  & \textbf{0.006}  & \textbf{0.005}  & 0.008  & 0.005  & 0.004  & 0.003  \\
    \midrule
    \multirow{4}[0]{*}{$\beta_1$} & DReER   & 0.017  & 0.014  & 0.012  & 0.009  & 0.025  & 0.020  & 0.015  & 0.014  \\
          & RDReER  & \textbf{0.001}  & \textbf{0.000}  & \textbf{0.000}  & \textbf{0.000}  & \textbf{0.002}  & \textbf{0.002}  & \textbf{0.001}  & \textbf{0.001}  \\
          & AReER   & 0.033  & 0.020  & 0.013  & 0.009  & 0.048  & 0.032  & 0.022  & 0.014  \\
          & RAReER  & 0.002  & 0.002  & 0.001  & \textbf{0.000}  & 0.004  & \textbf{0.002}  & \textbf{0.001}  & \textbf{0.001}  \\
    \midrule
    \multirow{4}[0]{*}{$\beta_2$} & DReER   & 0.015  & 0.010  & 0.008  & 0.006  & 0.033  & 0.016  & 0.015  & 0.010  \\
          & RDReER  & \textbf{0.001}  & \textbf{0.000}  & \textbf{0.000}  & \textbf{0.000}  & \textbf{0.002}  & \textbf{0.001}  & \textbf{0.001}  & \textbf{0.001}  \\
          & AReER   & 0.028  & 0.017  & 0.010  & 0.004  & 0.041  & 0.028  & 0.015  & 0.006  \\
          & RAReER  & 0.002  & 0.001  & 0.001  & \textbf{0.000}  & 0.003  & 0.002  & 0.002  & \textbf{0.001}  \\
    \midrule
    \multirow{4}[1]{*}{$\beta_3$} & DReER   & 0.016  & 0.012  & 0.012  & 0.009  & 0.027  & 0.016  & 0.015  & 0.012  \\
          & RDReER  & \textbf{0.001}  & \textbf{0.000}  & \textbf{0.000}  & \textbf{0.000}  & \textbf{0.002}  & \textbf{0.001} & \textbf{0.001}  & \textbf{0.00}1  \\
          & AReER   & 0.034  & 0.018  & 0.014  & 0.007  & 0.058  & 0.027  & 0.022  & 0.012  \\
          & RAReER  & 0.002  & 0.001  & 0.001  & \textbf{0.000}  & 0.004  & 0.002  & 0.002  & \textbf{0.001}  \\
    \midrule
    \multirow{3}[6]{*}{$\bbeta$} & \multirow{3}[6]{*}{Method} & \multicolumn{4}{c|}{Heterogeneous, $\e\sim \mathcal{N}(0,1)$} & \multicolumn{4}{c}{Heterogeneous, $\e\sim t(3)$} \\
\cmidrule{3-10}          &       & \multicolumn{8}{c}{$b$} \\
\cmidrule{3-10}          &       & 50    & 100   & 200   & 500   & 50    & 100   & 200   & 500 \\
    \midrule
    \multirow{4}[1]{*}{$\beta_0$} & DReER   & 0.453  & 0.434  & 0.418  & 0.424  & 1.363  & 1.179  & 1.182  & 1.119  \\
          & RDReER  & \textbf{0.004}  & \textbf{0.003}  & \textbf{0.002}  & \textbf{0.001}  & \textbf{0.006}  & \textbf{0.004}  & \textbf{0.003}  & \textbf{0.002}  \\
          & AReER   & 0.421  & 0.329  & 0.308  & 0.298  & 1.160  & 0.820  & 0.813  & 0.767  \\
          & RAReER  & 0.011  & 0.006  & 0.004  & 0.002  & 0.017  & 0.008  & 0.006  & 0.004  \\
    \midrule
    \multirow{4}[0]{*}{$\beta_1$} & DReER   & 0.025  & 0.021  & 0.017  & 0.013  & 0.049  & 0.036  & 0.025  & 0.021  \\
          & RDReER  & 0.006  & 0.006  & 0.005  & 0.005  & \textbf{0.009}  & 0.010  & 0.009  & 0.008  \\
          & AReER   & 0.044  & 0.028  & 0.017  & 0.012  & 0.101  & 0.066  & 0.043  & 0.025  \\
          & RAReER  & \textbf{0.005}  & \textbf{0.005}  & \textbf{0.003}  & \textbf{0.002}  & 0.011  & \textbf{0.007}  & \textbf{0.004}  & \textbf{0.004}  \\
    \midrule
    \multirow{4}[0]{*}{$\beta_2$} & DReER   & 0.027  & 0.021  & 0.018  & 0.016  & 0.082  & 0.037  & 0.035  & 0.025  \\
          & RDReER  & \textbf{0.003}  & \textbf{0.002} & \textbf{0.001} & \textbf{0.000}  & \textbf{0.006}  & \textbf{0.004}  & \textbf{0.004}  & 0.003  \\
          & AReER   & 0.037  & 0.023  & 0.013  & 0.008  & 0.090  & 0.056  & 0.033  & 0.015  \\
          & RAReER  & 0.010  & 0.007  & 0.005  & 0.003  & 0.012  & 0.006  & 0.006  & \textbf{0.002}  \\
    \midrule
    \multirow{4}[1]{*}{$\beta_3$} & DReER   & 0.019  & 0.012  & 0.012  & 0.007  & 0.046  & 0.024  & 0.019  & 0.010  \\
          & RDReER  & 0.021  & 0.022  & 0.021  & 0.020  & \textbf{0.015}  & 0.013  & 0.013  & 0.013  \\
          & AReER   & 0.040  & 0.018  & 0.015  & 0.006  & 0.110  & 0.049  & 0.036  & 0.016  \\
          & RAReER  & \textbf{0.014}  & \textbf{0.014}  & \textbf{0.013}  & \textbf{0.010}  & 0.016  & \textbf{0.011}  & \textbf{0.009}  & \textbf{0.007}  \\
    \bottomrule
    \end{tabular}%
  \label{tab:robust_out}%
    \end{table}%
    
    \item[2)] Under fixed $N_b$ scenarios, increasing the batch size $n_t$ leads to lower MSEs for both DReER and AReER, likely due to the improved efficiency of the LM statistic. When $n_t$ is fixed, increasing the number of batches $b$ further reduces the MSEs of both methods, with their performance approaching that of the Oracle estimator. This provides empirical evidence supporting the consistency of the proposed procedures.
    
    \item[3)] DReER outperforms AReER when the detection mechanism is effective. In such cases, DReER achieves estimation accuracy comparable to that of the Oracle benchmark. However, as the number of abnormal batches increases, particularly under heterogeneous models or heavy-tailed error distributions, the performance of DReER deteriorates and becomes more sensitive to the batch size.
    
    \item[4)] When heterogeneity is marginal or evolves gradually, AReER demonstrates superior performance. By down-weighting potentially heterogeneous batches rather than excluding them entirely, AReER mitigates the risk of incorrect detection and yields more stable estimation.
    
    \item[5)] When the error terms follow a heavy-tailed distribution, the overall estimation accuracy declines relative to the standard normal-error scenario. Incorporating a robust loss function substantially improves performance, leading to lower MSEs and reducing the dependence on batch size. In addition, the robust approach effectively handles outliers, as illustrated in Section~\ref{sub:robust}.
\end{itemize}

\section{Empirical Study}\label{sec:empirical}
% To illustrate the proposed methods, we analyze the Parkinson’s Telemonitoring dataset from the UCI Machine Learning Repository (\href{https://archive.ics.uci.edu/dataset/189/parkinsons+telemonitoring}{Parkinsons Telemonitoring Dataset}). The dataset aims to predict the total Unified Parkinson’s Disease Rating Scale (UPDRS) score based on voice measurements. It contains 5,875 observations from $42$ individuals with early-stage Parkinson’s disease and includes 16 biomedical voice features. Monitoring the progression of Parkinson’s disease typically relies on time-consuming clinical assessments conducted by trained professionals, making large-scale and individualized modeling challenging in real-world settings \citep{Parkinson2010Tsanas}. Meanwhile, substantial patient-level heterogeneity arises from differences in disease progression and physiological characteristics, implying that data from different patients or batches may follow distinct underlying mechanisms. Ignoring such heterogeneity may lead to biased estimation and degraded predictive performance. Therefore, it is necessary to incorporate detection or adaptive renewable procedures to dynamically accommodate heterogeneity across data batches.

To illustrate the proposed methods, we analyze the Parkinson's Telemonitoring dataset from the UCI Machine Learning Repository (\href{https://archive.ics.uci.edu/dataset/189/parkinsons+telemonitoring}{Parkinsons Telemonitoring Dataset}). It contains 5,875 observations from 42 individuals with early-stage Parkinson's disease and 16 biomedical voice features for predicting the total Unified Parkinson's Disease Rating Scale (UPDRS) score. Because clinical assessment of disease progression is time-consuming and requires trained professionals \citep{Parkinson2010Tsanas}, voice-based monitoring offers a scalable alternative for tracking disease severity. However, patient-level heterogeneity in disease progression, physiological characteristics, and recording conditions may induce batch-specific data-generating mechanisms, motivating detection-based and adaptive renewable procedures.

The remote data-collection scheme also makes this dataset suitable for a streaming analysis. Since observations were recorded at patients' homes and transmitted through the Internet, they can be viewed as sequentially arriving records. We pooled observations from the first 10 days across all patients to form the initial batch. The remaining streaming data were then processed sequentially by patient ID, with each patient forming one online batch. Under this setup, the training data consist of 43 batches, and patient-specific batches contain roughly 100--135 observations each. For prediction, observations in the last \(H \in \{3,5,7,10\}\)-days were reserved as the testing set. 
%Because the observation times are irregular, the testing sample size varies with $H$. 
Predictive performance was evaluated on the pooled testing set using the expectile loss (EL),
\begin{align*}
\mathrm{EL}_H(\tau)
=
\frac{1}{n_H}
\sum_{i=1}^{n_H}
\rho_{\tau}\{y_i^{\mathrm{test}}-\hat{y}_i^{\mathrm{test}}(\tau)\},
\end{align*}
where $n_H$ is the number of pooled testing observations under horizon $H$, and $y_i^{\mathrm{test}}$ and $\hat{y}_i^{\mathrm{test}}(\tau)$ denote the observed and predicted responses in the testing set, respectively, here the testing set is pooled across patients.

\begin{figure}[htbp]
    \centering
    \includegraphics[width=.8\linewidth]{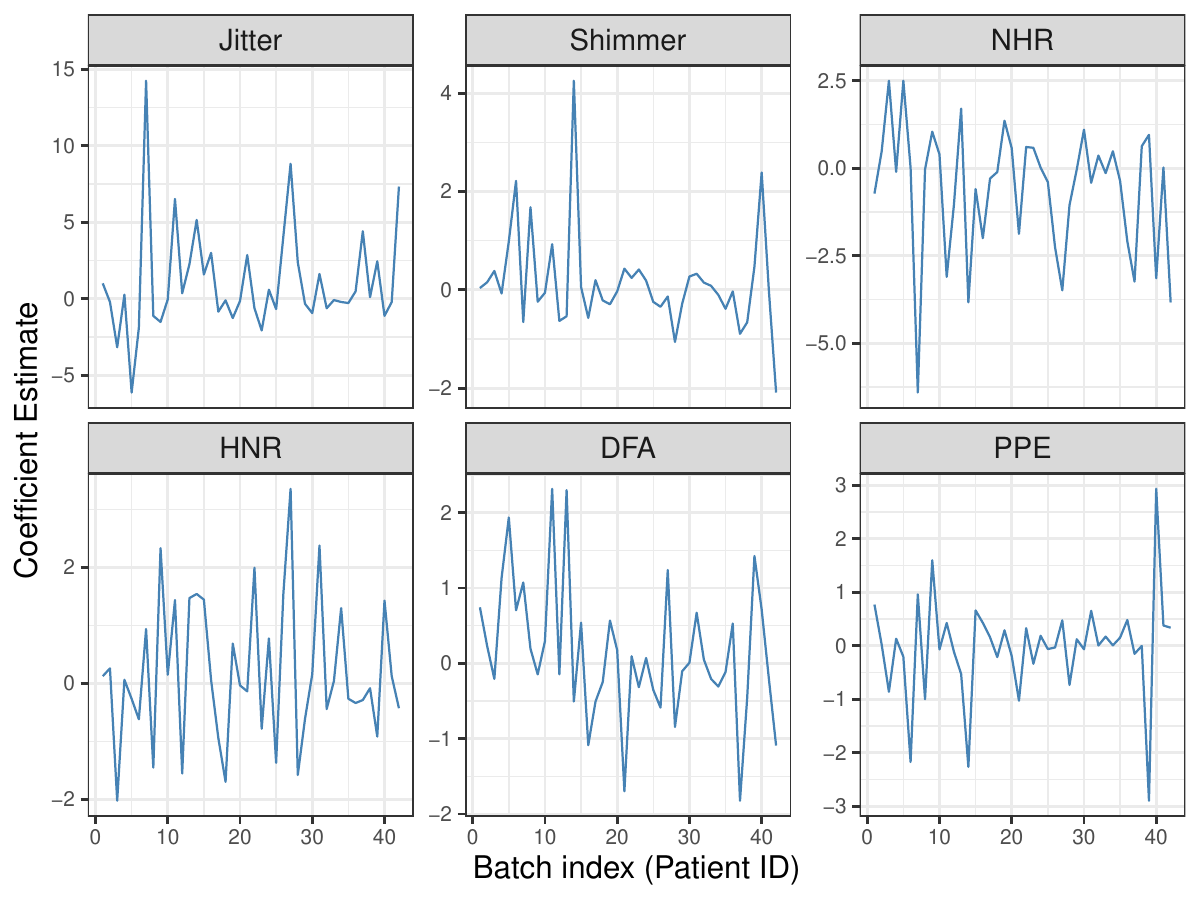}
    \caption{The batch-level estimates at $\tau = 0.5$.}
    \label{fig:batch_est}
\end{figure}

According to \citet{Parkinson2010Tsanas}, six covariates were identified as significant predictors. Therefore, we consider these variables, namely Jitter, Shimmer, NHR, HNR, DFA, and PPE, as the explanatory variables in our analysis. Figure~\ref{fig:batch_est} presents the batch-level coefficient estimates for the Parkinson’s dataset, revealing substantial heterogeneity across data batches, as indicated by the considerable variability in the coefficient estimates. Table~\ref{tab:empirical} reports the estimated coefficients at expectile levels $\tau \in \{0.1,\ldots,0.9\}$. In most cases, the estimated signs are consistent across methods, except for Jitter and NHR, for which DReER and RDReER yield signs that differ from those obtained by the other approaches. However, as shown in Figure~\ref{fig:batch_est}, the average coefficient estimates at $\tau=0.5$ for Jitter and NHR are positive and negative, respectively, which is consistent with the estimates obtained from DReER and RDReER. This suggests that the detection-based methods may better capture the dominant structural patterns across batches by effectively filtering out anomalous batches. 

\begin{figure}[htbp]
    \centering
    \includegraphics[width=1\linewidth]{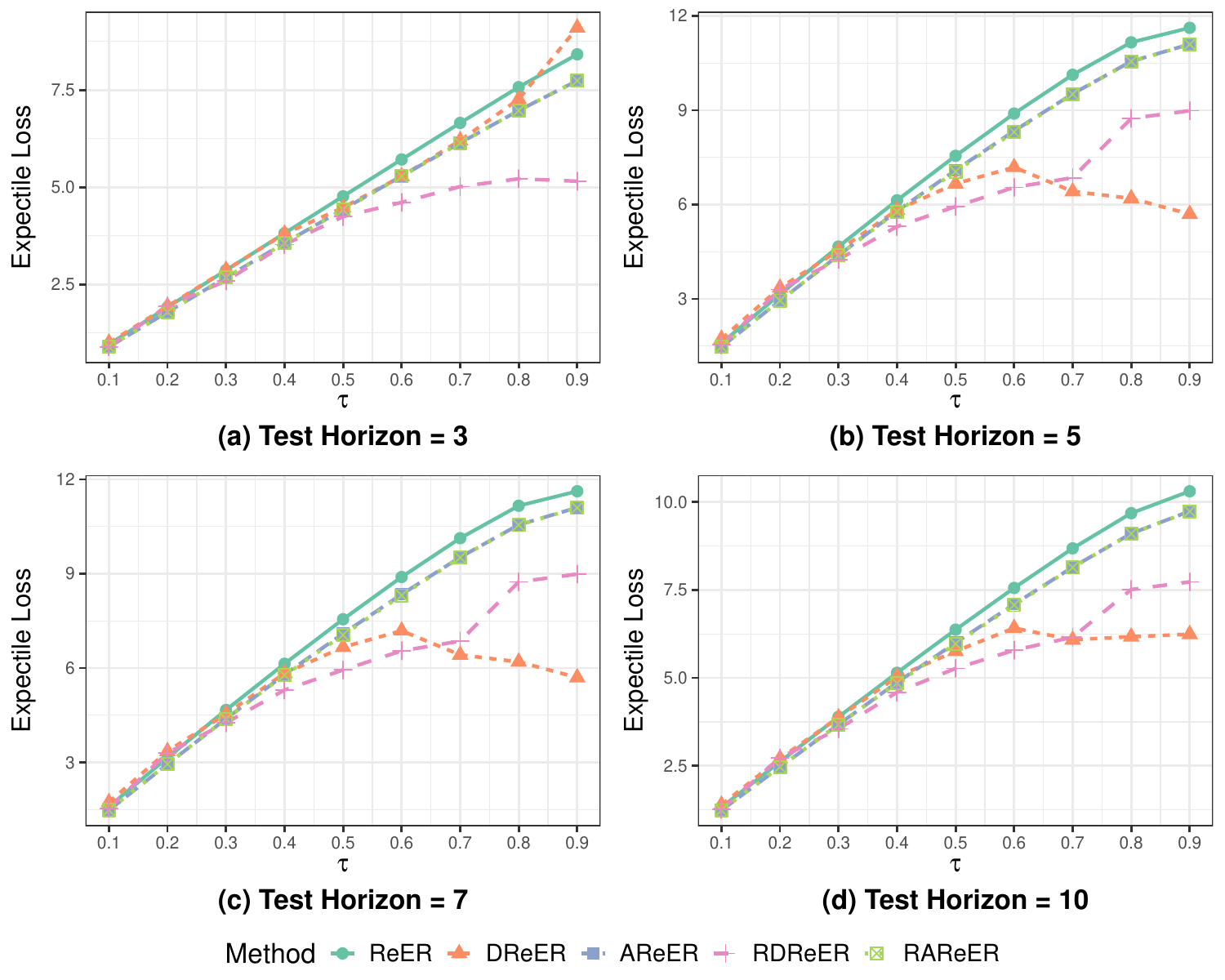}
    \caption{Expectile loss under different testing window sizes $H \in \{3,5,7,10\}$ at $\tau = 0.1,\ldots,0.9$.}
    \label{fig:EL}
\end{figure}

Figure~\ref{fig:EL} displays the EL values of the renewable methods under different forecast horizons. Both the detection-based and adaptive-weighting approaches consistently achieve lower prediction losses than ReER. Moreover, as the forecast horizon $H$ increases, the detection-based method exhibits superior predictive accuracy. This finding suggests that batch-level heterogeneity has a more pronounced impact on longer-term prediction. By explicitly identifying and excluding potentially abnormal batches, the detection-based method adopts a more conservative estimation strategy, thereby improving predictive stability.

\section{Conclusion and Discussion}\label{sec:conclusion}

In this paper, we develop a renewable expectile regression framework to address batch-level heterogeneity in streaming data settings. Building upon an LM-based testing procedure, we propose two complementary strategies: a detection-based method (DReER), which identifies and excludes heterogeneous batches from subsequent renewable updates, and an adaptive method (AReER), which mitigates the impact of heterogeneity by down-weighting potentially abnormal batches. Furthermore, we extend the framework to a robust renewable procedure by replacing the quadratic loss with the Huber loss, thereby improving stability under heavy-tailed errors and in the presence of outliers.

Extensive simulation studies demonstrate that renewable methods explicitly accounting for batch-level heterogeneity can effectively mitigate the bias induced by abnormal batches and achieve performance comparable to that of the Oracle benchmark. The two strategies exhibit complementary strengths: the detection-based approach performs better when heterogeneity is pronounced and can be reliably detected, whereas the adaptive approach is preferable when heterogeneity is mild or evolves gradually. Incorporating the robust loss further improves performance under heavy-tailed error distributions, reducing sensitivity to extreme observations and enhancing overall estimation stability. Finally, the real-data application demonstrates that the proposed methods substantially outperform the naive renewable procedure that indiscriminately aggregates all incoming batches.

To conclude, we acknowledge several limitations of the present work and outline directions for future research. First, the proposed framework assumes that all data batches are generated from a common underlying model with a fixed but unknown parameter. As such, it is not explicitly designed to accommodate model drift, in which the underlying data-generating mechanism evolves over time, as discussed in \citet{chen_renewable_2024}. Nevertheless, the adaptive strategy provides a natural starting point for addressing this issue. For instance, constructing a modified response $\breve{y}_{ti}$ that combines the renewable estimator with batch-specific estimators may allow the procedure to adapt more flexibly to structural changes across batches.

Second, the proposed renewable estimation framework primarily uses the accumulated Hessian matrix as a compressed summary of historical data information for subsequent online updating. However, from a strict definitional perspective, the accumulated Hessian matrix in expectile regression involves asymmetric weights that depend on the estimated parameters. To accommodate the streaming data setting, the current study approximates these weights using the online updated estimates from the corresponding historical batches, rather than recalculating them using the latest parameter estimates. This approximation may affect the convergence properties and, potentially, the convergence rate of the resulting estimator. Future research may consider incorporating a dynamic reconstruction strategy, in which the stored summary information from historical batches is combined with a local Taylor expansion to correct the resulting approximation error \citep{Han04062026}.

Lastly, the current methodology primarily focuses on low-dimensional settings. In modern large-scale applications, however, extremely high-dimensional data are increasingly prevalent. Extending the renewable framework to incorporate penalty-based regularization or subsampling techniques \citep{pan2021distributed, li2024poisson, chen2024estimation} would therefore be a promising and practically important direction. Such extensions could further enhance the scalability and applicability of renewable expectile regression in high-dimensional streaming environments.

%\wss{please update some literature. some without number or pages. }

\section*{Acknowledgments}
	
This research was financially supported by the State Key Program of National Natural Science Foundation of China [Nos. 72531002].
	
\bibliography{mybib}
\bibliographystyle{apalike}

\appendix
\setcounter{table}{0}
\setcounter{figure}{0}
\section{Technical Proofs} \label{app:proof}

\subsection{Adaptive renewable estimation}\label{areer-loss}

Define the adaptive loss for each batch as $\loss_{n_t}(\bbeta)=\gamma_t\cdot\loss_{n_t}(\bbeta)+(1-\gamma_t)\cdot\ploss_{n_t}(\bbeta)$, 
where $\ploss_{n_t}(\bbeta)=\frac{1}{n_t}\rho_{\tau}\left(\breve{y}_{ti}-\x_{ti}^\top\bbeta\right)$ and $\breve{y}_{ti}=\x_{ti}^\top\truebeta$.
And the overall adaptive loss up to batch $b$ is
$$\renewloss_{N_b}(\bbeta)=\frac{1}{N_b}\sum_{t=1}^tn_t\cdot\renewloss_{n_t}(\bbeta).$$

For the first batch $\D_1$, the adaptive renewable estimator is:
$$\renew_1=\mathop{\arg\min}_{\bbeta}\renewloss_{n_1}(\bbeta)=\mathop{\arg\min}_{\bbeta}\left\{\gamma_1\cdot\loss_{n_1}(\bbeta)+(1-\gamma_1)\cdot\ploss_{n_1}(\bbeta)\right\},$$

When the second batch $\D_2$ arrives the overall loss function is:
$$\renewloss_{N_2}(\bbeta)=\frac{1}{N_2}\left[n_1\renewloss_{n_1}(\bbeta)+n_2\renewloss_{n_2}(\bbeta)\right],$$
which can be approximated by the following loss function (after take Taylor expansion of $\renewloss_{n_1}(\bbeta)$ around $\renew_1$):
$$\renewloss_{N_2}(\bbeta)=\frac{1}{N_2}\left\{\frac{n_1}{2}\left(\bbeta-\renew_1\right)^\top\left[\gamma_1\Hess_{n_1}(\renew_1)+(1-\gamma_1)\pHess_{n_1}(\renew_1)\right]\left(\bbeta-\renew_1\right)+n_2\renewloss_{n_2}(\bbeta)\right\}.$$

When the third batch $\D_3$ arrives the loss function is:
$$\renewloss_{N_3}(\bbeta)=\frac{1}{N_3}\left[n_1\renewloss_{n_1}(\bbeta)+n_2\renewloss_{n_2}(\bbeta)+n_3\renewloss_{n_3}(\bbeta)\right]=\frac{1}{N_3}\left[N_2\renewloss_{N_2}(\bbeta)+n_3\renewloss_{n_3}(\bbeta)\right],$$
and which can be approximately as:
\begin{align*}
\begin{aligned}
\renewloss_{N_3}(\bbeta)
= \frac{1}{N_3}\Bigg\{ 
&\frac{n_1}{2}(\bbeta-\renew_2)^\top
\left[ \gamma_1\Hess_{n_1}(\renew_1)
      +(1-\gamma_1)\pHess_{n_1}(\renew_1)
\right]
(\bbeta-\renew_2) + \\
&\frac{n_2}{2}(\bbeta-\renew_2)^\top
\left[ \gamma_2\Hess_{n_2}(\renew_2)
      +(1-\gamma_2)\pHess_{n_2}(\renew_2)
\right]
(\bbeta-\renew_2) +  n_3\renewloss_{n_3}(\bbeta)
\Bigg\}  \\
=\frac{1}{N_3}\Bigg\{ 
&\frac{1}{2}(\bbeta-\renew_2)^\top
\sum_{t=1}^2\left(\gamma_t\cdot n_t\cdot\Hess_{n_t}(\renew_t)
      +(1-\gamma_t)\cdot n_t\cdot\pHess_{n_t}(\renew_t)\right)
(\bbeta-\renew_2) +\\ & n_3\times\left[\gamma_3\cdot\loss_{n_3}(\bbeta)+(1-\gamma_3)\cdot\ploss_{n_3}(\bbeta)\right]\Bigg\}.
\end{aligned}
\end{align*}

Finally, generalizing the procedure to streaming datasets up to batch $b$, the renewable loss can be defined as:
\begin{align*}
\begin{aligned} 
\renewloss_{\Nb}(\bbeta) =\frac{1}{\Nb}\Bigg\{&\frac{1}{2}(\bbeta-\renew_{b-1})^\top\sumb\left(\gamma_t\cdot\nt \cdot\Hess_{\nt}(\renew_{t})+(1-\gamma_t)\cdot\nt \cdot\pHess_{\nt}(\renew_{t})\right)(\bbeta-\renew_{b-1})+\\
&n_b\times\left[\gamma_b\cdot\loss_{\nb}(\bbeta)+(1-\gamma_b)\cdot\ploss_{\nb}(\bbeta)\right]\Bigg\}.
\end{aligned}
\end{align*}

\subsection{Technical proofs of Proposition~\ref{P1}.} \label{proof-sub3}

According to the renewable estimation formula of AReER, the estimator based on the first $b$ data batches is given by
\begin{equation}
\tilde{\boldsymbol{\beta}}_b^a
=
\left[
\mathbf H_{N_{b-1}}^a
+
\mathbf W_{n_b}^{a}
\right]^{-1}
\left[
\mathbf H_{N_{b-1}}^a
\tilde{\boldsymbol{\beta}}_{b-1}^{a}
+
\mathbf U_{n_b}^{a}
\right],
\label{eq:areer-proof-start}
\end{equation}
where
$\mathbf W_{n_b}^{a}
=\gamma_b
\mathbf W_{n_b}(\tilde{\boldsymbol{\beta}}_{b-1}^{a})
+(1-\gamma_b)
\breve{\mathbf W}_{n_b}(\tilde{\boldsymbol{\beta}}_{b-1}^{a})$,
$\mathbf U_{n_b}^{a}
=\gamma_b\mathbf U_{n_b}(\tilde{\boldsymbol{\beta}}_{b-1}^{a})
+(1-\gamma_b)
\breve{\mathbf U}_{n_b}(\tilde{\boldsymbol{\beta}}_{b-1}^{a})$, and
\[
\mathbf H_{N_{b-1}}^a
=
\sum_{t=1}^{b-1}
\left[
\gamma_t\mathbf W_{n_t}(\tilde{\boldsymbol{\beta}}_t^a)
+
(1-\gamma_t)\breve{\mathbf W}_{n_t}(\tilde{\boldsymbol{\beta}}_t^a)
\right].
\]

Under the regularity conditions,
$\tilde{\boldsymbol{\beta}}_t^a$ and
$\tilde{\boldsymbol{\beta}}_{t-1}^a$
lie in a neighborhood of $\boldsymbol{\beta}_0$. Therefore, the standard expansion gives
\begin{equation}
\mathbf H_{N_{b-1}}^a
=
\sum_{t=1}^{b-1}
\left[
\gamma_t
\{n_t\mathbf A+o_p(n_t)\}
+
(1-\gamma_t)
\{n_t\mathbf A+o_p(n_t)\}
\right]
=
N_{b-1}\mathbf A+o_p(N_{b-1}).
\label{eq:H-expand}
\end{equation}
Similarly,
$\mathbf W_{n_b}^{a}=n_b\mathbf A+o_p(n_b)$.
Combining this result with Eq.\eqref{eq:H-expand}, we obtain
\begin{equation}
\mathbf H_{N_{b-1}}^a+\mathbf W_{n_b}^{a}
=
N_b\mathbf A+o_p(N_b).
\label{eq:denominator-expand}
\end{equation}

By assumption, the true parameter for batch $b$ is defined as
$\boldsymbol{\beta}_b=\boldsymbol{\beta}_0+\boldsymbol{\eta}_b$,
which satisfies the first-order condition $\mathbb E
\left\{
\nabla \mathcal L_{n_b}(\boldsymbol{\beta}_b)
\right\}
=\mathbf 0$. 
Therefore, for any point in a neighborhood of $\boldsymbol{\beta}_0$, we have
\begin{equation*}
\mathbf U_{n_b}(\bar{\boldsymbol{\beta}})
=
n_b\mathbf A
(\boldsymbol{\beta}_0+\boldsymbol{\eta}_b)
+
o_p(n_b).
\end{equation*}
Taking
$\bar{\boldsymbol{\beta}}=\tilde{\boldsymbol{\beta}}_{b-1}^{a}$ gives
\begin{equation}
\mathbf U_{n_b}(\tilde{\boldsymbol{\beta}}_{b-1}^{a})
=
n_b\mathbf A
(\boldsymbol{\beta}_0+\boldsymbol{\eta}_b)
+
o_p(n_b).
\label{eq:U-real-expand}
\end{equation}
For the pseudo-response variable
$\breve y_{bi}
=\boldsymbol{x}_{bi}^{\top}\tilde{\boldsymbol{\beta}}_{b-1}^{a}$,
which is anchored at $\tilde{\boldsymbol{\beta}}_{b-1}^{a}$, we have
\begin{equation}
\breve{\mathbf U}_{n_b}(\tilde{\boldsymbol{\beta}}_{b-1}^{a})
=
n_b\mathbf A
\tilde{\boldsymbol{\beta}}_{b-1}^{a}
+
o_p(n_b).
\label{eq:U-guided-expand}
\end{equation}
Combining Eqs.\eqref{eq:U-real-expand} and\eqref{eq:U-guided-expand}, we obtain
\begin{equation}
\begin{aligned}
\mathbf U_{n_b}^{a}
&=
\gamma_b
\mathbf U_{n_b}(\tilde{\boldsymbol{\beta}}_{b-1}^{a})
+
(1-\gamma_b)
\breve{\mathbf U}_{n_b}(\tilde{\boldsymbol{\beta}}_{b-1}^{a})
\\
&=
\gamma_b n_b\mathbf A
(\boldsymbol{\beta}_0+\boldsymbol{\eta}_b)
+
(1-\gamma_b)n_b\mathbf A
\tilde{\boldsymbol{\beta}}_{b-1}^{a}
+
o_p(n_b)
\\
&=
n_b\mathbf A
\left[
\boldsymbol{\beta}_0
+
\gamma_b\boldsymbol{\eta}_b
+
(1-\gamma_b)
\left(
\tilde{\boldsymbol{\beta}}_{b-1}^{a}
-
\boldsymbol{\beta}_0
\right)
\right]
+
o_p(n_b).
\end{aligned}
\label{eq:Ua-expand}
\end{equation}

Substituting Eqs.\eqref{eq:denominator-expand} and\eqref{eq:Ua-expand}
into Eq.\eqref{eq:areer-proof-start}, we obtain
\begin{equation}
\begin{aligned}
\tilde{\boldsymbol{\beta}}_b^a
&=
\left\{
N_b\mathbf A+o_p(N_b)
\right\}^{-1}
\Bigg[
\left\{
N_{b-1}\mathbf A+o_p(N_{b-1})
\right\}
\tilde{\boldsymbol{\beta}}_{b-1}^{a}
\\
&\qquad\qquad
+n_b\mathbf A
\Big[
\boldsymbol{\beta}_0
+\gamma_b\boldsymbol{\eta}_b
+(1-\gamma_b)
\big(
\tilde{\boldsymbol{\beta}}_{b-1}^{a}
-\boldsymbol{\beta}_0
\big)
\Big]
+o_p(n_b)
\Bigg].
\end{aligned}
\label{eq:substitute-expand}
\end{equation}
Define
$\mathbf d_t
=\tilde{\boldsymbol{\beta}}_t^a-\boldsymbol{\beta}_0$.
Then, Eq.\eqref{eq:substitute-expand} implies $\tilde{\boldsymbol{\beta}}_b^a
=
\boldsymbol{\beta}_0
+
\frac{N_{b-1}}{N_b}\mathbf d_{b-1}
+
\frac{n_b}{N_b}
\left[
\gamma_b\boldsymbol{\eta}_b
+
(1-\gamma_b)\mathbf d_{b-1}
\right]
+
o_p(n_b/N_b)$.
% \begin{equation*}
% \tilde{\boldsymbol{\beta}}_b^a
% =
% \boldsymbol{\beta}_0
% +
% \frac{N_{b-1}}{N_b}\mathbf d_{b-1}
% +
% \frac{n_b}{N_b}
% \left[
% \gamma_b\boldsymbol{\eta}_b
% +
% (1-\gamma_b)\mathbf d_{b-1}
% \right]
% +
% o_p(1).
% \end{equation*}
Subtracting $\boldsymbol{\beta}_0$ from both sides gives
\begin{equation}
\mathbf d_b
=
\frac{N_{b-1}}{N_b}\mathbf d_{b-1}
+
\frac{n_b}{N_b}
\left[
\gamma_b\boldsymbol{\eta}_b
+
(1-\gamma_b)\mathbf d_{b-1}
\right]
+
o_p(n_b/N_b).
\label{eq:single-step-d}
\end{equation}
Multiplying both sides by $N_b$, Eq.\eqref{eq:single-step-d} becomes
\begin{equation}\label{eq:eq:single-step-d2}
N_b\mathbf d_b
=
N_{b-1}\mathbf d_{b-1}
+
n_b
\left[
\gamma_b\boldsymbol{\eta}_b
+
(1-\gamma_b)\mathbf d_{b-1}
\right]
+
o_p(n_b).
\end{equation}
Assume $N_0=0$ and that the cumulative remainder satisfies
$\sum_{t=1}^b o_p(n_t)=o_p(N_b)$. Iterating Eq.\eqref{eq:eq:single-step-d2} over
$t=1,\ldots,b$ yields
$N_b\mathbf d_b
=
\sum_{t=1}^b
n_t
\left[
\gamma_t\boldsymbol{\eta}_t
+
(1-\gamma_t)\mathbf d_{t-1}
\right]
+
o_p(N_b)$.
Dividing both sides by $N_b$ and taking expectations, we have
\begin{equation}
\mathbb E
\left(
\tilde{\boldsymbol{\beta}}_b^a-\boldsymbol{\beta}_0
\right)
=
\frac1{N_b}
\sum_{t=1}^b
n_t
\left[
\gamma_t\boldsymbol{\eta}_t
+
(1-\gamma_t)
\mathbb E
\left(
\tilde{\boldsymbol{\beta}}_{t-1}^{a}
-
\boldsymbol{\beta}_0
\right)
\right]
+
o(1).
\label{eq:bias-with-history}
\end{equation}
By the assumption of Proposition~\ref{P1}, the contribution from the guiding center satisfies
\[
{N_b}^{-1}
\sum_{t=1}^b
n_t(1-\gamma_t)
\mathbb E
\left(
\tilde{\boldsymbol{\beta}}_{t-1}^{a}
-
\boldsymbol{\beta}_0
\right)
=
o(1).
\]
Furthermore, define
\[
\bar{\boldsymbol{\eta}}_b
=
\frac{
\sum_{t=1}^b n_t\gamma_t\boldsymbol{\eta}_t
}{
\sum_{t=1}^b n_t\gamma_t
}.
\]
Substituting this definition into Eq.\eqref{eq:bias-with-history}, we obtain
\begin{equation*}
\mathbb E
\left(
\tilde{\boldsymbol{\beta}}_b^a-\boldsymbol{\beta}_0
\right)
=
\left(
\sum_{t=1}^b
\frac{n_t}{N_b}
\gamma_t
\right)
\bar{\boldsymbol{\eta}}_b
+
o(1).
\end{equation*}
\section{Additional simulation results}

\subsection{Additional results of \ref{simu:case2}: Block abnormal batches}

\begin{figure}[htbp]
    \centering
    \includegraphics[width=1\linewidth]{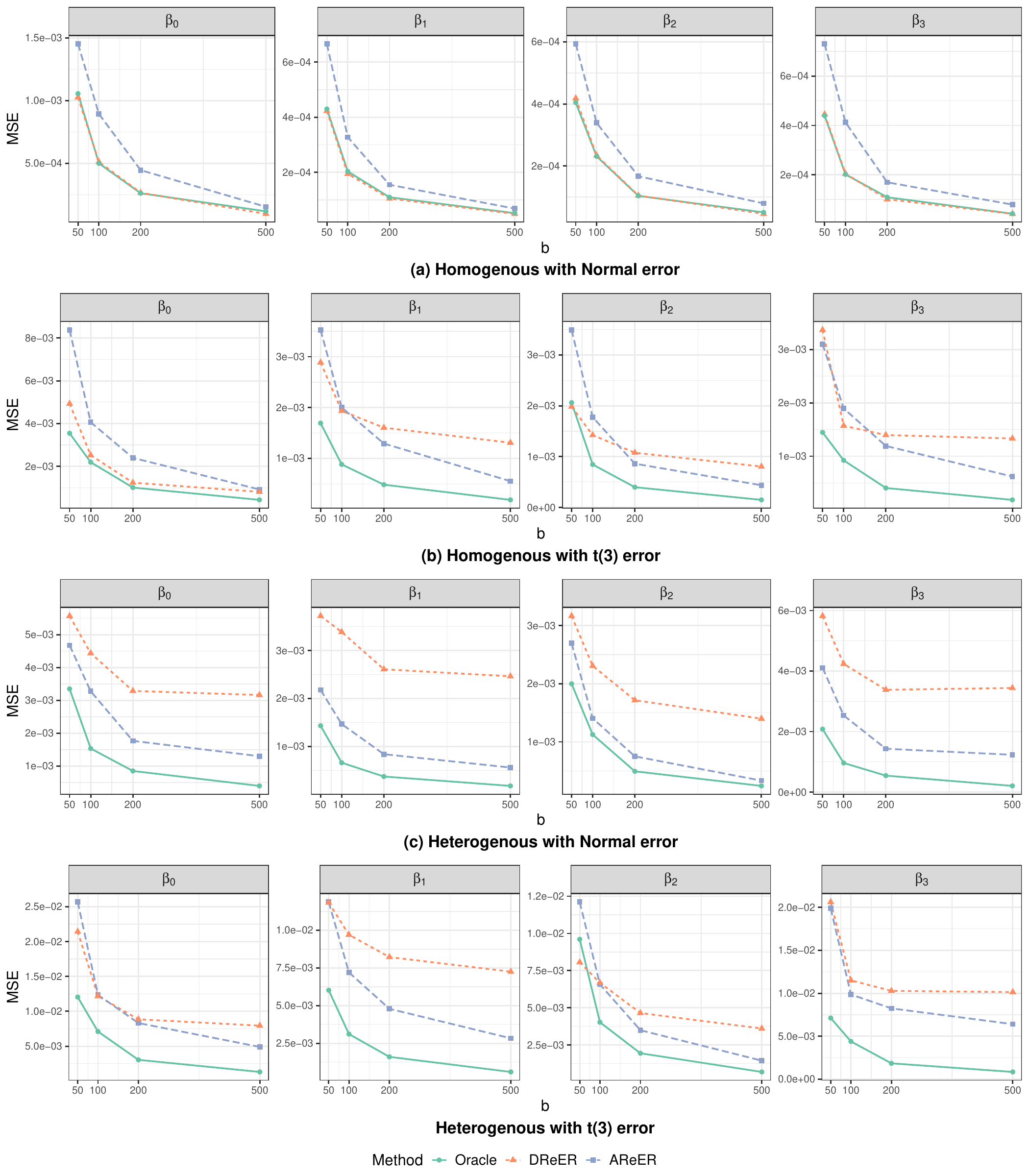}
    \caption{MSE values under fixed total sample size $n_t=200$ with varying batch size $b \in \{50,100,200,500\}$. All abnormal batches appear within the initial 30\% of the data stream.}
    \label{fig:MSE_block_30}
\end{figure}

\begin{figure}[htbp]
    \centering
    \includegraphics[width=1\linewidth]{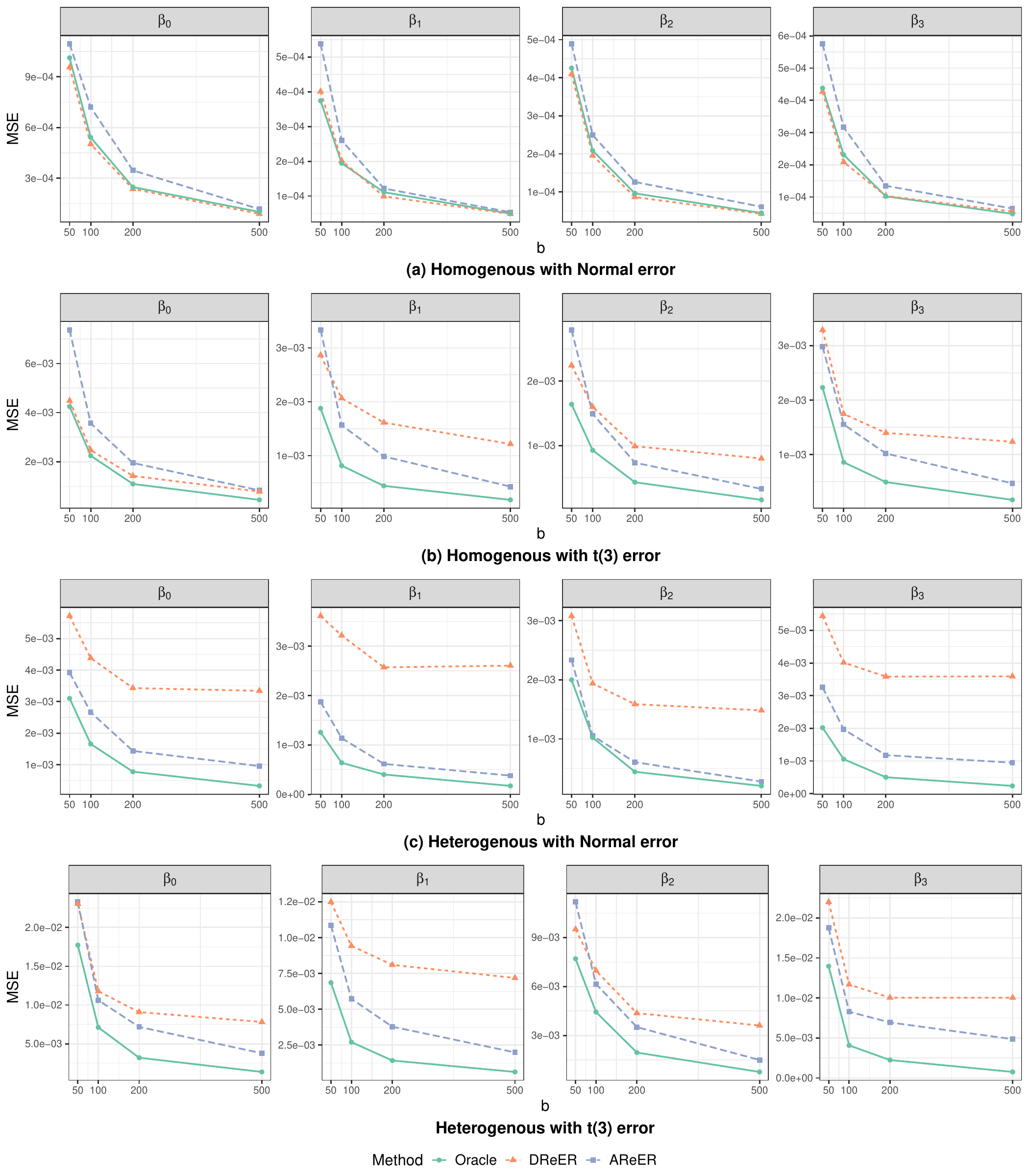}
    \caption{MSE values under fixed total sample size $n_t=200$ with varying batch size $b \in \{50,100,200,500\}$. All abnormal batches appear within the last 70\% of the data stream.}
    \label{fig:MSE_block_70}
\end{figure}

Tables~\ref{tab:block30} and~\ref{tab:block70} present MSE values under a fixed total sample size $N_b = 100,000$, with batch sizes $n_t \in \{100, 200, 500, 1000\}$, where abnormal batches occur in the initial 30\% and final 70\% of the data stream, respectively. Figures~\ref{fig:MSE_block_30} and~\ref{fig:MSE_block_70} show the MSEs for a fixed batch size $n_t = 200$ and varying numbers of batches $ b \in \{50, 100, 200, 500\}$ under the same two abnormal-batch settings.

\begin{table}[htbp]
  \centering
  \renewcommand{\arraystretch}{0.8}
  \renewcommand\tabcolsep{9.0pt}
  \caption{MSE for $\tau=0.25$ with fixed $N_b=100{,}000$, and varying $n_t \in \{100, 200, 500, 1000\}$. The abnormal batch appear in the first 30\%.}
  \label{tab:block30}
  \small
  \begin{tabular}{cc|cccc|cccc}
  \toprule
  \multirow{3}[6]{*}{$\boldsymbol{\beta}$} 
  & \multirow{3}[6]{*}{Method} 
  & \multicolumn{8}{c}{$n_t$} \\
  \cmidrule{3-10}
  & & 200 & 500 & 1000 & 2000 & 200 & 500 & 1000 & 2000 \\
  \cmidrule{3-10}
  & & \multicolumn{4}{c|}{Homogeneous, $\e \sim \mathcal{N}(0,1)$} 
  & \multicolumn{4}{c}{Homogeneous, $\e \sim t(3)$} \\
  \midrule
  $\beta_0$ & Oracle & 0.011 & 0.010 & 0.010 & 0.009 & 0.039 & 0.036 & 0.038 & 0.049 \\
            & DReER  & 0.009 & 0.009 & 0.010 & 0.009 & 0.080 & 0.036 & 0.038 & 0.052 \\
            & AReER  & 0.018 & 0.017 & 0.017 & 0.015 & 0.092 & 0.068 & 0.080 & 0.103 \\
  \addlinespace
  $\beta_1$ & Oracle & 0.005 & 0.004 & 0.005 & 0.004 & 0.020 & 0.022 & 0.014 & 0.019 \\
            & DReER  & 0.005 & 0.004 & 0.005 & 0.004 & 0.137 & 0.025 & 0.016 & 0.019 \\
            & AReER  & 0.008 & 0.007 & 0.008 & 0.008 & 0.069 & 0.038 & 0.033 & 0.034 \\
  \addlinespace
  $\beta_2$ & Oracle & 0.005 & 0.004 & 0.004 & 0.004 & 0.018 & 0.021 & 0.019 & 0.019 \\
            & DReER  & 0.005 & 0.004 & 0.004 & 0.004 & 0.073 & 0.022 & 0.018 & 0.019 \\
            & AReER  & 0.008 & 0.006 & 0.007 & 0.007 & 0.040 & 0.036 & 0.043 & 0.034 \\
  \addlinespace
  $\beta_3$ & Oracle & 0.005 & 0.004 & 0.006 & 0.004 & 0.021 & 0.019 & 0.020 & 0.019 \\
            & DReER  & 0.004 & 0.004 & 0.005 & 0.004 & 0.131 & 0.020 & 0.020 & 0.020 \\
            & AReER  & 0.008 & 0.010 & 0.007 & 0.006 & 0.062 & 0.035 & 0.029 & 0.044 \\
  \midrule
  \multicolumn{2}{c|}{} 
  & \multicolumn{4}{c|}{Heterogeneous, $\e \sim \mathcal{N}(0,1)$} 
  & \multicolumn{4}{c}{Heterogeneous, $\e \sim t(3)$} \\
  \midrule
  $\beta_0$ & Oracle & 0.035 & 0.030 & 0.034 & 0.050 & 0.142 & 0.117 & 0.115 & 0.170 \\
            & DReER  & 0.320 & 0.034 & 0.032 & 0.028 & 0.828 & 0.459 & 0.207 & 0.170 \\
            & AReER  & 0.140 & 0.058 & 0.050 & 0.053 & 0.526 & 0.298 & 0.294 & 0.353 \\
  \addlinespace
  $\beta_1$ & Oracle & 0.016 & 0.013 & 0.016 & 0.025 & 0.071 & 0.064 & 0.048 & 0.064 \\
            & DReER  & 0.257 & 0.017 & 0.014 & 0.014 & 0.767 & 0.340 & 0.120 & 0.065 \\
            & AReER  & 0.066 & 0.026 & 0.025 & 0.027 & 0.350 & 0.155 & 0.126 & 0.119 \\
  \addlinespace
  $\beta_2$ & Oracle & 0.023 & 0.019 & 0.021 & 0.034 & 0.088 & 0.101 & 0.088 & 0.090 \\
            & DReER  & 0.143 & 0.019 & 0.020 & 0.019 & 0.357 & 0.198 & 0.094 & 0.090 \\
            & AReER  & 0.032 & 0.033 & 0.034 & 0.035 & 0.137 & 0.141 & 0.210 & 0.168 \\
  \addlinespace
  $\beta_3$ & Oracle & 0.023 & 0.021 & 0.027 & 0.035 & 0.098 & 0.089 & 0.094 & 0.091 \\
            & DReER  & 0.330 & 0.028 & 0.025 & 0.019 & 1.063 & 0.511 & 0.190 & 0.099 \\
            & AReER  & 0.123 & 0.050 & 0.035 & 0.031 & 0.657 & 0.281 & 0.165 & 0.205 \\
  \bottomrule
  \end{tabular}%
  \begin{tablenotes}
    \footnotesize
    \item Metrics are reported in units of $10^{-2}$.
  \end{tablenotes}
\end{table}

\begin{table}[htbp]
  \centering
  \renewcommand{\arraystretch}{0.8}
  \renewcommand\tabcolsep{9.0pt}
  \caption{MSE for $\tau=0.25$ with fixed $N_b=100{,}000$, and varying $n_t \in \{100, 200, 500, 1000\}$. The abnormal batch appear in the last 70\%.}
  \label{tab:block70}
  \small
  \begin{tabular}{cc|cccc|cccc}
  \toprule
  \multirow{3}[6]{*}{$\boldsymbol{\beta}$} 
  & \multirow{3}[6]{*}{Method} 
  & \multicolumn{8}{c}{$n_t$} \\
  \cmidrule{3-10}
  & & 200 & 500 & 1000 & 2000 & 200 & 500 & 1000 & 2000 \\
  \cmidrule{3-10}
  & & \multicolumn{4}{c|}{Homogeneous, $\e \sim \mathcal{N}(0,1)$} 
  & \multicolumn{4}{c}{Homogeneous, $\e \sim t(3)$} \\
  \midrule
  $\beta_0$ & Oracle & 0.013 & 0.011 & 0.010 & 0.010 & 0.040 & 0.043 & 0.038 & 0.040 \\
            & DReER  & 0.012 & 0.010 & 0.011 & 0.010 & 0.087 & 0.044 & 0.038 & 0.041 \\
            & AReER  & 0.014 & 0.014 & 0.014 & 0.013 & 0.076 & 0.055 & 0.063 & 0.072 \\
  \addlinespace
  $\beta_1$ & Oracle & 0.004 & 0.005 & 0.005 & 0.004 & 0.020 & 0.017 & 0.016 & 0.015 \\
            & DReER  & 0.004 & 0.005 & 0.005 & 0.004 & 0.146 & 0.022 & 0.017 & 0.015 \\
            & AReER  & 0.006 & 0.007 & 0.006 & 0.006 & 0.053 & 0.029 & 0.025 & 0.024 \\
  \addlinespace
  $\beta_2$ & Oracle & 0.006 & 0.004 & 0.004 & 0.005 & 0.019 & 0.022 & 0.018 & 0.018 \\
            & DReER  & 0.005 & 0.004 & 0.004 & 0.004 & 0.075 & 0.023 & 0.015 & 0.019 \\
            & AReER  & 0.007 & 0.005 & 0.005 & 0.005 & 0.030 & 0.028 & 0.032 & 0.025 \\
  \addlinespace
  $\beta_3$ & Oracle & 0.005 & 0.005 & 0.006 & 0.005 & 0.018 & 0.017 & 0.018 & 0.017 \\
            & DReER  & 0.005 & 0.004 & 0.005 & 0.004 & 0.136 & 0.020 & 0.016 & 0.016 \\
            & AReER  & 0.007 & 0.007 & 0.006 & 0.005 & 0.045 & 0.027 & 0.023 & 0.032 \\
  \midrule
  \multicolumn{2}{c|}{} 
  & \multicolumn{4}{c|}{Heterogeneous, $\e \sim \mathcal{N}(0,1)$} 
  & \multicolumn{4}{c}{Heterogeneous, $\e \sim t(3)$} \\
  \midrule
  $\beta_0$ & Oracle & 0.041 & 0.036 & 0.034 & 0.036 & 0.131 & 0.139 & 0.113 & 0.137 \\
            & DReER  & 0.323 & 0.037 & 0.033 & 0.034 & 0.847 & 0.457 & 0.192 & 0.145 \\
            & AReER  & 0.110 & 0.049 & 0.043 & 0.046 & 0.407 & 0.240 & 0.227 & 0.262 \\
  \addlinespace
  $\beta_1$ & Oracle & 0.013 & 0.017 & 0.016 & 0.015 & 0.071 & 0.057 & 0.056 & 0.053 \\
            & DReER  & 0.253 & 0.019 & 0.014 & 0.014 & 0.771 & 0.362 & 0.118 & 0.053 \\
            & AReER  & 0.044 & 0.025 & 0.021 & 0.021 & 0.251 & 0.122 & 0.097 & 0.085 \\
  \addlinespace
  $\beta_2$ & Oracle & 0.027 & 0.021 & 0.019 & 0.022 & 0.088 & 0.104 & 0.088 & 0.087 \\
            & DReER  & 0.142 & 0.020 & 0.018 & 0.020 & 0.350 & 0.209 & 0.087 & 0.089 \\
            & AReER  & 0.027 & 0.026 & 0.027 & 0.027 & 0.144 & 0.120 & 0.174 & 0.131 \\
  \addlinespace
  $\beta_3$ & Oracle & 0.024 & 0.022 & 0.027 & 0.023 & 0.089 & 0.076 & 0.079 & 0.081 \\
            & DReER  & 0.337 & 0.027 & 0.024 & 0.022 & 1.084 & 0.497 & 0.160 & 0.085 \\
            & AReER  & 0.094 & 0.037 & 0.028 & 0.027 & 0.508 & 0.223 & 0.127 & 0.155 \\
  \bottomrule
  \end{tabular}%
  \begin{tablenotes}
    \footnotesize
    \item Metrics are reported in units of $10^{-2}$.
  \end{tablenotes}
\end{table}

\subsection{Additional results of of Robust Renewable ER based Method}
\begin{figure}[htbp]
    \centering
    \includegraphics[width=.845\linewidth]{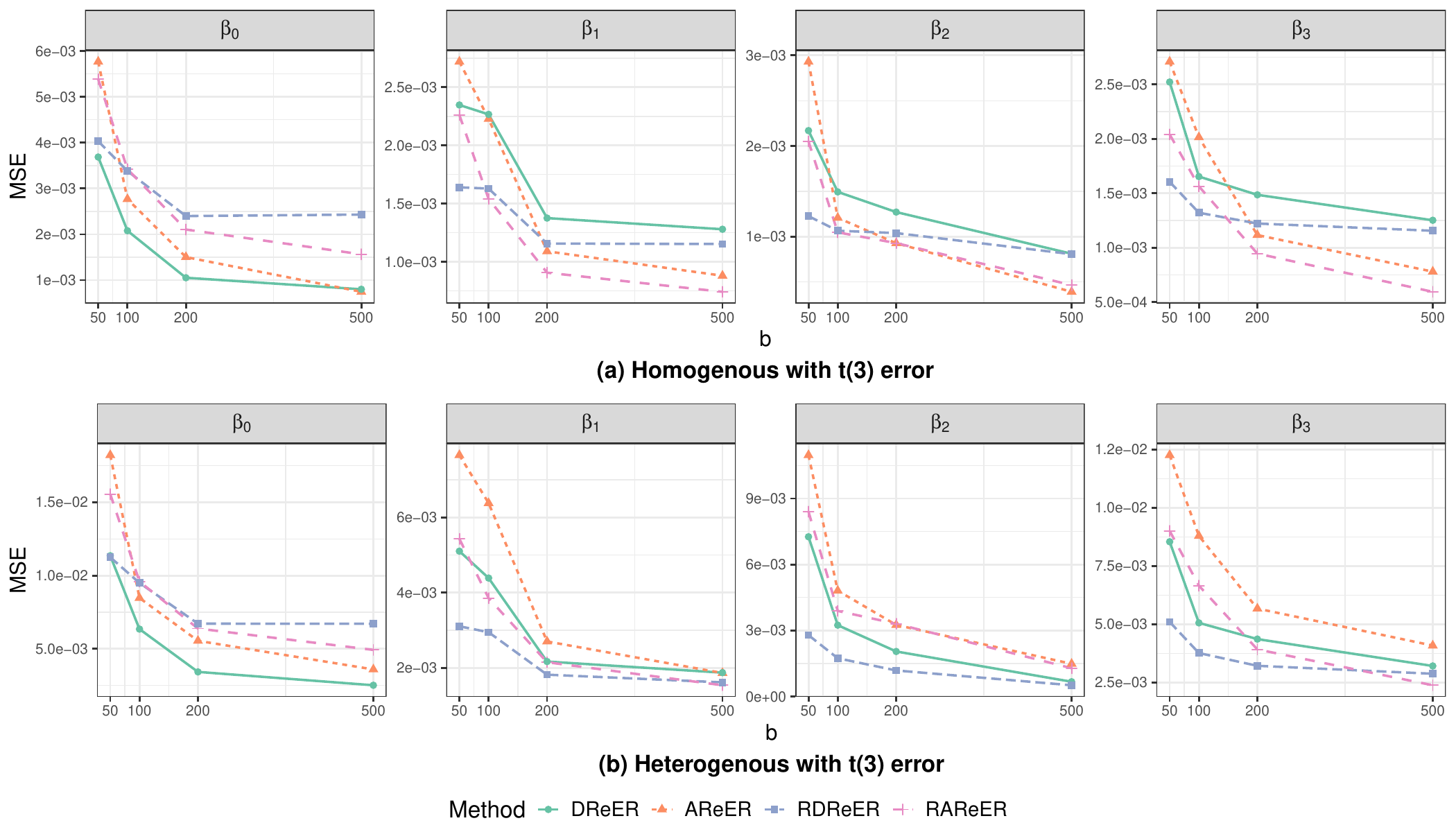}
    \caption{MSEs of robust renewable methods with fixed batch size $n_t = 200$ and varying numbers of batches $b \in \{50, 100, 200, 500\}$ under \ref{simu:case3}, with an abnormal rate of 30\%.}
    \label{fig:MSE_small_R}
\end{figure}
\begin{figure}[htbp]
    \centering
    \includegraphics[width=.845\linewidth]{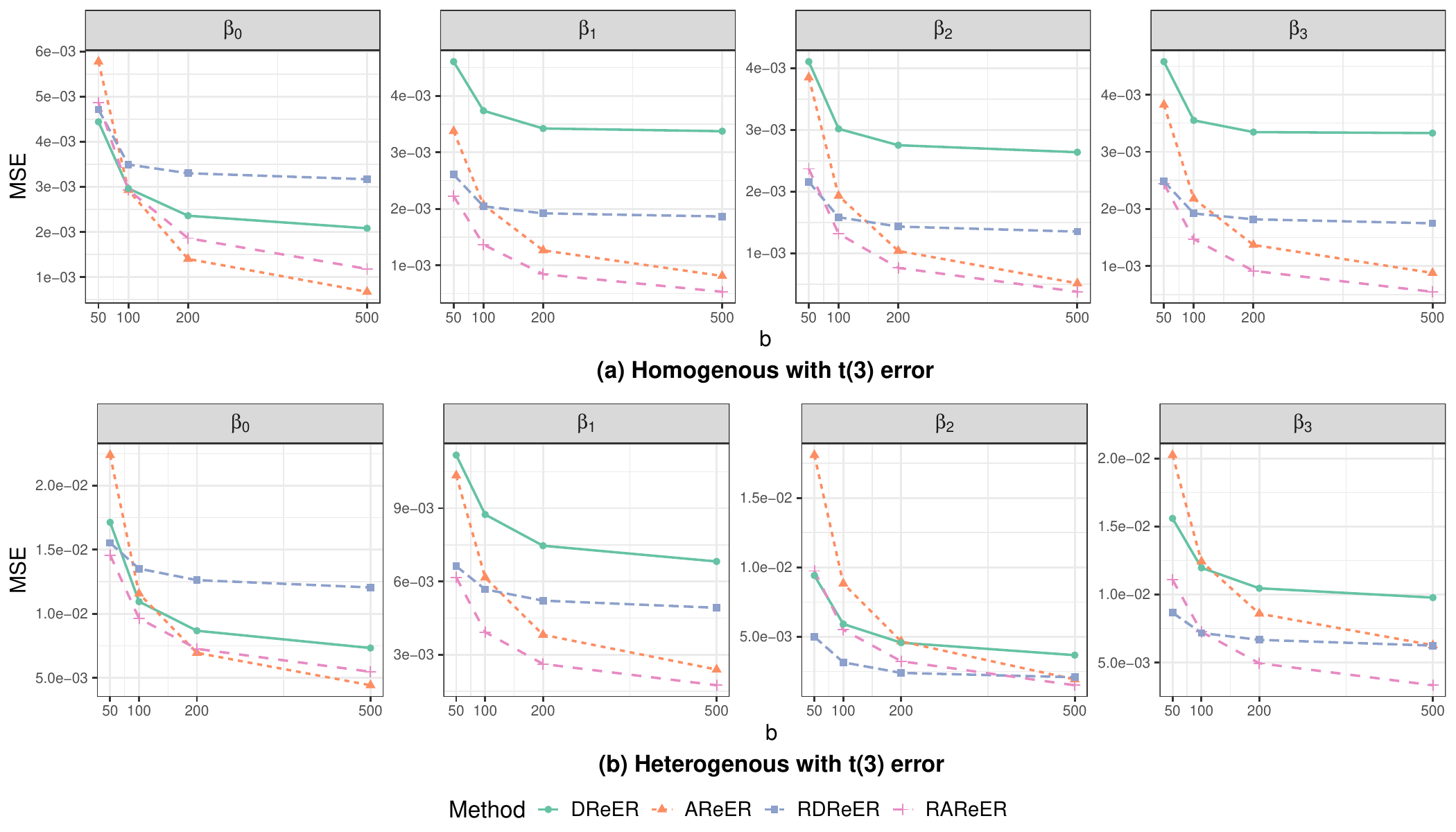}
    \caption{MSEs of robust renewable methods with fixed batch size $n_t = 200$ and varying numbers of batches $b \in \{50, 100, 200, 500\}$ under \ref{simu:case4}.}
    \label{fig:MSE_drift_R}
\end{figure}

\newpage
\subsection{Additional results of empirical study}

\begin{table}[htbp]
 \small
  \renewcommand{\arraystretch}{0.7}
  \setlength{\tabcolsep}{6pt}
  \centering
  \caption{The coefficients estimation of Parkinson dataset at $\tau = 0.1,\ldots,0.9$.}
    \begin{tabular}{ccccccccccc}
    \toprule
    \multirow{2}[4]{*}{Variable} & \multirow{2}[4]{*}{Method} & \multicolumn{9}{c}{$\tau$} \\
\cmidrule{3-11}          &       & 0.1   & 0.2   & 0.3   & 0.4   & 0.5   & 0.6   & 0.7   & 0.8   & 0.9 \\
    \midrule
    \multirow{5}[2]{*}{Jitter} & ReER  & -1.119  & -1.172  & -1.231  & -1.300  & -1.376  & -1.461  & -1.565  & -1.708  & -1.992  \\
          & DReER   & -0.091  & 0.393  & 0.978  & 0.899  & 0.536  & 0.285  & 0.053  & -0.196  & -3.044  \\
          & AReER    & -1.619  & -1.625  & -1.639  & -1.661  & -1.703  & -1.750  & -1.814  & -1.918  & -2.207  \\
          & RDReER  & -1.136  & -0.530  & -0.028  & 0.333  & 0.164  & -0.394  & -0.615  & 2.385  & 3.226  \\
          & RAReER  & -1.590  & -1.603  & -1.613  & -1.627  & -1.678  & -1.742  & -1.814  & -1.919  & -2.209  \\
    \midrule
    \multirow{5}[2]{*}{Shimmer} & ReER  & 1.064  & 1.034  & 1.000  & 0.963  & 0.920  & 0.868  & 0.800  & 0.697  & 0.479  \\
          & DReER    & 0.245  & 0.246  & 0.827  & 1.336  & 1.233  & 1.344  & 0.815  & 0.106  & -1.349  \\
          & AReER    & 0.561  & 0.543  & 0.517  & 0.486  & 0.421  & 0.355  & 0.280  & 0.173  & -0.064  \\
          & RDReER  & 0.077  & 0.657  & 0.187  & 0.638  & 0.632  & 0.273  & 0.296  & -0.461  & -0.520  \\
          & RAReER  & 0.546  & 0.536  & 0.513  & 0.485  & 0.421  & 0.355  & 0.279  & 0.172  & -0.064  \\
    \midrule
    \multirow{5}[2]{*}{HNR} & ReER  & 1.869  & 1.791  & 1.707  & 1.613  & 1.509  & 1.392  & 1.254  & 1.079  & 0.801  \\
          & DReER    & 2.165  & 1.979  & 1.159  & 1.188  & 1.245  & 0.930  & 0.898  & 0.637  & -0.984  \\
          & AReER    & 1.409  & 1.376  & 1.332  & 1.284  & 1.175  & 1.086  & 0.998  & 0.880  & 0.664  \\
          & RDReER  & 1.546  & 1.622  & 1.292  & 0.953  & 0.724  & 0.666  & 0.510  & -0.029  & -0.254  \\
          & RAReER  & 1.414  & 1.378  & 1.335  & 1.284  & 1.170  & 1.082  & 0.996  & 0.879  & 0.663  \\
    \midrule
    \multirow{5}[2]{*}{NHR} & ReER  & 1.331  & 1.359  & 1.394  & 1.435  & 1.479  & 1.526  & 1.580  & 1.650  & 1.780  \\
          & DReER    & 1.487  & 1.029  & -0.058  & -0.652  & -0.726  & -1.144  & -1.188  & -0.835  & -0.117  \\
          & AReER    & 1.778  & 1.778  & 1.784  & 1.797  & 1.807  & 1.831  & 1.871  & 1.929  & 2.110  \\
          & RDReER  & 1.805  & 1.275  & 0.831  & 0.040  & -0.246  & 0.217  & -0.067  & -2.723  & -4.067  \\
          & RAReER  & 1.779  & 1.773  & 1.770  & 1.768  & 1.777  & 1.816  & 1.868  & 1.929  & 2.110  \\
    \midrule
    \multirow{5}[2]{*}{DFA} & ReER  & -0.586  & -0.595  & -0.604  & -0.613  & -0.623  & -0.632  & -0.637  & -0.631  & -0.579  \\
          & DReER    & -0.729  & -0.767  & -0.819  & -0.598  & -0.422  & -0.264  & 0.389  & 0.758  & 1.389  \\
          & AReER    & -0.730  & -0.738  & -0.745  & -0.748  & -0.756  & -0.752  & -0.742  & -0.727  & -0.661  \\
          & RDReER  & -0.835  & -0.899  & -0.724  & -0.649  & -0.502  & -0.404  & -0.275  & -1.359  & -1.500  \\
          & RAReER  & -0.732  & -0.739  & -0.742  & -0.736  & -0.740  & -0.743  & -0.739  & -0.726  & -0.660  \\
    \midrule
    \multirow{5}[2]{*}{PPE} & ReER  & 0.098  & 0.070  & 0.040  & 0.007  & -0.027  & -0.064  & -0.101  & -0.138  & -0.164  \\
          & DReER    & 0.970  & 0.750  & -0.244  & -0.408  & -0.309  & -0.443  & 0.068  & 0.312  & 0.535  \\
          & AReER    & -0.233  & -0.232  & -0.236  & -0.240  & -0.267  & -0.278  & -0.277  & -0.267  & -0.228  \\
          & RDReER  & 0.246  & 0.242  & -0.104  & -0.426  & -0.549  & -0.772  & -0.980  & -0.334  & -0.443  \\
          & RAReER  & -0.221  & -0.227  & -0.231  & -0.239  & -0.271  & -0.281  & -0.278  & -0.268  & -0.228  \\
    \bottomrule
    \end{tabular}%
  \label{tab:empirical}%
\end{table}%

\end{document}